\documentclass[preprint]{JHEP3} 

\JHEPspecialurl{http://jhep.sissa.it/JOURNAL/JHEP3.tar.gz}

\usepackage{epsfig,multicol}

\def \arctanh{\mathop{\rm arctanh}\nolimits}
\def \arccoth{\mathop{\rm arccoth}\nolimits}

\def \arcsinh{\mathop{\rm arcsinh}\nolimits}

\newcommand\fverb{\setbox\fverbbox=\hbox\bgroup\verb}
\newcommand\fverbdo{\egroup\medskip\noindent%
			\fbox{\unhbox\fverbbox}\ }
\newcommand\fverbit{\egroup\item[\fbox{\unhbox\fverbbox}]}
\newbox\fverbbox

\title{On domain walls in a Ginzburg-Landau non-linear ${\mathbb{S}}^2$-sigma model}

\author{A. Alonso Izquierdo$^{a,c}$, M.A. Gonz\'alez Le\'on$^{a,c}$, J. Mateos Guilarte$^{b,c}$ and M. de la Torre Mayado$^{b,c}$ \\
$^{a}$ Departamento de Matematica
Aplicada, Universidad de Salamanca, SPAIN
\\$^{b}$ Departamento de Fisica Fundamental, Universidad de Salamanca, SPAIN
\\$^{c}$ IUFFyM, Universidad de Salamanca, SPAIN
}

\abstract{ The domain wall solutions of a Ginzburg-Landau non-linear ${\mathbb S}^2$-sigma hybrid model
are unveiled. There are three types of basic topological walls and two types of degenerate families of composite
- one topological, the other non-topological- walls. The domain wall solutions are identified as the finite action trajectories (in infinite time) of a related mechanical system that is Hamilton-Jacobi separable in sphero-conical coordinates. The physical and mathematical features of these domain walls are thoroughly discussed.}


\begin{document}


\section{Introduction}

Domain walls are two-dimensional membranes that form when a discrete symmetry is broken at a phase transition, e.g.,
the interfaces (Bloch, Ising walls) between magnetic domains in ferromagnetic materials. In Cosmological models of the early Universe domain walls form according to a pattern known as the (second) Kibble mechanism; see \cite{Kibble}. The
impact of domain wall defects and of other topological defects in Cosmology has been studied in depth in the monograph
\cite{VilShel}. The evolution of domain wall networks is a problem of particular interest in this context, see e.g. \cite{OMA}. This problem is usually studied in computer simulations, although an analytic approach has been developed in
\cite{ENS} and \cite{ENS1}. By identifying the moduli space of domain wall networks in a $U(N_C)$ gauge theory with
$N_f$ scalar fields in the adjoint representation, the authors of \cite{ENS} and \cite{ENS1} implemented the low-energy dynamics of the network by studying the geodesic motion in the domain wall network moduli space. It is also interesting to consider domain walls as the seed of Randall-Sundrum scenarios \cite{RS}, in which space-time is five-dimensional and the 3-brane is wrapped around some background four-dimensional gravitational field, while the particle dynamics is concentrated inside the wall; see \cite{BST}-\cite{CDL}-\cite{BG}-\cite{BBL}-\cite{SAH}. In this framework some authors have considered the possibility that the Big Bang of the standard (3+1) dimensional cosmology was originated from the collision of two branes within a higher dimensional spacetime, leading to the production of a large
amount of entropy \cite{Blanco}.

Our purpose here is to investigate the very rich moduli space of domain walls in a hybrid of the non-linear sigma model and the Ginzburg-Landau theory of phase transitions. The linear ${\mathbb O}(N)$-sigma model
is the key to our present understanding about the origin of mass: see \cite{Gell-Mann} and \cite{Velt}. The non-linear ${\mathbb S}^N$-model version only describes the dynamics of Goldstone bosons. In \cite{AMAJ}, we addressed
an especially simple case of massive non-linear sigma model: we chose ${\mathbb S}^2$ as the target manifold; assigned
different masses to the two pseudo-Goldstone bosons, and were able to identify all the domain walls of the system

Regarding \cite{AMAJ}, topological defects {\footnote{ Of varying character, depending on the spatial dimension and the charge: kinks, Q-kinks, lumps, strings, walls, etcetera .}}  in massive non-linear sigma
models have been known for some time and have been profusely studied in
different supersymmetric models under the circumstance that all
masses of the pseudo-Nambu-Goldstone particles are equal. The study
started with two papers by Abraham and Townsend \cite{AT},
\cite{AT1}, in which the authors discovered a family of Q-kinks in a
(1+1)-dimensional ${\cal N}=(4,4)$ supersymmetric non-linear sigma
model with a hyper-Kahler Gibbons-Hawking instanton as the target
space and mass terms obtained from dimensional reduction. In
\cite{Nit}, however, these kinks were re-considered by constructing
the dimensionally reduced supersymmetric model by means of the
mathematically elegant technique of hyper-Kahler quotients. By doing
so, the authors dealt with massive ${\mathbb CP}^N$  or ${\mathbb
HP}^N$ models: a playground closer to our simpler massive ${\mathbb
S}^2$-sigma model. Similar ${\cal N}=2$ BPS walls in the ${\mathbb
CP}^1$-model with twisted mass were described in \cite{Dor}. In a
parallel development in the (2+1)-dimensional version of these
models, two-dimensional Q-lumps were discovered in \cite{Lee} and
\cite{Abr}. Within this field, the most interesting result is
the demonstration in \cite{GPTT} and \cite{INOS} that composite
solitons in ${\it d}=3+1$ of Q-strings and domain walls are exact
BPS solutions that preserve $\frac{1}{4}$ of the supersymmetries: (
See also the review \cite{EINOS}, where a summary of these
supersymmetric topological solitons is offered.)

Our research differed from the above works in two important
aspects: 1) We stuck to a purely bosonic framework. 2) We studied
the case when the masses of the pseudo Nambu-Goldstone bosons are
different, a property that forbids extended supersymmetries. The search for
domain walls in the ${\it d}=3+1$-model is tantamount to the search for
finite action trajectories in the repulsive Neumann system \cite{Neumann}: a
particle moving in an ${\mathbb S}^2$-sphere under the action of
non-isotropic repulsive elastic forces. It is well known that this
dynamical system is completely integrable \cite{Moser}, \cite{Dubrovin}. We
showed, however, that the problem is Hamilton-Jacobi separable by
using elliptic coordinates in the sphere. Use of this property allowed us to
find four families of homoclinic trajectories starting and ending at
one of the poles which are unstable points of the mechanical system.
In the field-theoretical model, the poles become ground states,
whereas the homoclinic trajectories correspond to four families of
non-topological domain walls. Each member in a family is formed by a
non-linear combination of two basic topological domain walls (of different
type), with their centers located at any relative distance with
respect each other.

Here we shall address a Ginzburg-Landau non-linear ${\mathbb S}^2$-sigma model; i.e.,
we will keep the target space but we add a quartic, rather than quadratic, independent of field gradients,
contribution to the potential energy density; see \cite{DF} for a mathematical definition of these models.
Because of the constraint, the GL function must be non-isotropic and we shall consider non-equal quadratic and quartic
couplings in such a way that the anisotropy is maximal. The consequence is the existence of a spontaneously broken discrete symmetry: a necessary condition for the existence of domain walls.

We shall further restrict (but not too much) the space of parameters of the model, the quadratic and quartic couplings
besides the radius of the ${\mathbb S}^2$-sphere, bearing in mind that we have to deal with an integrable analogous
mechanical problem. Instead  of the Neummann system we must solve the problem of a particle moving on the ${\mathbb S}^2$-sphere under the action of non-isotropic inelastic (non-harmonic) forces. Fortunately, this mechanical system is also Hamilton-Jacobi separable, and we shall apply the Hamilton-Jacobi procedure to find all the finite mechanical action trajectories by using sphero-conical coordinates. There are four unstable points -four ground states in the field theory-; three types of basic heteroclinic trajectories -topological domain walls- joining unstable points; four families of heteroclinic trajectories joining antipodal unstable points -topological domain walls- with the same mechanical action -wall tension-, and four families of homoclinic trajectories -non-topological domain walls- also with the same mechanical action{\footnote{The wall tensions of the topological and non-topological families are different.}}. We remark that the basic domain walls are usual walls, concentrated at a point. Domain walls belonging to any of the degenerate families are composite domain walls in the sense that, generically, the walls are centered at two points, resembling  a non-linear superposition of two basic walls.

In order to describe all this, we shall organize the paper as follows: In Section \S. 2 we introduce the model, explain the physical content, and describe the basic domain walls that can be found by applying the trial orbit method. Section \S. 3 is devoted to addressing the analogous mechanical problem. Sphero-conical coordinates are used to show the
Hamilton-Jacobi separability. A new basic wall is easily guessed, and the other two, previously known, are also
expressed in these coordinates. In Section \S. 4 we apply the Hamilton-Jacobi procedure in full generality. We find
the families of non-topological and topological domain walls in an explicit analytic form. This is remarkable: the solution for the orbit and time-schedule equations provided by the HJ prescription is frequently expressed in an implicit form that is difficult to invert. We have succeeded, however, in performing the inversion in this problem. We offer a last Section \S. 5 with further comments and some suggestions for future lines of enquiry.

Finally we have complemented this paper with a MATHEMATICA file, which can be found at http://campus.usal.es/$\sim$mpg/General/Mathematicatools.htm. This file includes animated figures which display the behaviour of the domain wall families depending on the different coupling constants and family parameters.

\clearpage
\section{The Ginzburg-Landau non-linear ${\mathbb S}^2$-sigma model}

The action and the constraint governing the dynamics of this hybrid of Ginzburg-Landau and non-linear Sigma models are respectively:
\begin{eqnarray*}
&&S[\vec{\chi}]=\int \, d^4y \, \left\{\frac{1}{2}\frac{\partial\vec{\chi}}{\partial y^\mu}\cdot\frac{\partial\vec{\chi}}{\partial y_\mu}-\frac{1}{2} \left(\sum_{a=1}^3\alpha_a^2\chi_a^2-m^2\right)^2-\frac{1}{2}
\sum_{b=1}^3\beta_b^2\chi_b^2\right\}\\ &&\vec{\chi}(y^\mu)\cdot \vec{\chi}(y^\mu)=\chi_1^2(y^\mu)+
\chi_2^2(y^\mu)+\chi_3^2(y^\mu)=m^2 R^2  \quad .
\end{eqnarray*}
 Owing to the constraint, the fields take values in the ${\mathbb S}^2$-sphere of radius $m R$ embedded in ${\mathbb R}^3$. Setting an ortho-normal frame, $\vec{e}_a\cdot\vec{e}_b=\delta_{ab}$, $a,b=1,2,3$, in ${\mathbb R}^3$, we write
the fields in the form:
\[
\vec{\chi}\, : \, {\mathbb R}^{1,3} \, \longrightarrow \, \, {\mathbb S}^2\quad;\qquad \vec{\chi}(y^\mu)=\sum_{a=1}^3 \, \chi_a(y^\mu)\vec{e}_a \, \quad .
\]
 $\vec{\chi}(y^\mu)$ are maps from the Minkowski space ${\mathbb R}^{1,3}$ into ${\mathbb S}^2\subset {\mathbb R}^3$. The contra-variant tetra-vector $y^\mu$, $\mu=0,1,2,3$, provides
local coordinates for a point in ${\mathbb R}^{1,3}$: the Minkowski space equipped with the metric tensor
$g^{\mu\nu}={\rm diag}(1,-1,-1,-1)$. Thus, $y^\mu y_\mu=y_0^2-y_1^2-y_2^2-y_3^2$, $\frac{\partial}{\partial y^\mu}\frac{\partial}{\partial y_\mu}=\frac{\partial^2}{\partial y_0^2}-\frac{\partial^2}{\partial y_1^2}
-\frac{\partial^2}{\partial y_2^2}-\frac{\partial^2}{\partial y_3^2}$, etcetera.

 Throughout the paper we shall use the natural system of units, in which the Planck constant and the speed of light in vacuum are in the units: $\hbar=c=1$. Therefore, the physical dimensions of the fields and the parameters $m$ and $\beta_a$ are those of inverse length, $[\chi_a]=[\beta_a]=[m]=L^{-1}$, whereas $\alpha_a$ and $R$ are non-dimensional couplings. In terms of  non-dimensional fields $\phi_a=\frac{1}{m}\chi_a$, space-time coordinates $x^\mu=m y^\mu$, and quadratic couplings
$\eta_a^2=\frac{1}{m^2}\beta_a^2$, the action and the constraint read{\footnote{In $V(\phi_1,\phi_2,\phi_3)$ we have dropped the irrelevant constant: $R^2\left(\frac{\eta_3^2-\eta_1^2}{2}+(1-\alpha_3^2R^2)(\alpha_1^2-\alpha_3^2)\right)-\frac{\eta_3^2 R^2}{2}$.}}:
\begin{eqnarray*}
S[\vec{\phi}]&=& \int d^4 x \left\{ \frac{1}{2}
\partial_\mu\vec{\phi}\cdot \partial^\mu \vec{\phi}
 -V(\phi_1,\phi_2,\phi_3)\right\} \, \, \, , \quad \phi_1^2(x^\mu)+\phi_2^2(x^\mu)+\phi_3^2(x^\mu)=R^2 \nonumber
\\
&& V(\phi_1,\phi_2,\phi_3)=\frac{1}{2} \left( \sum_{a=1}^3 \alpha_a^2
\phi_a^2 - 1 \right)^2+\frac{1}{2} \sum_{b=1}^3 \eta_b^2
\phi_b^2 \quad .\label{action2}
\end{eqnarray*}
We shall address the maximally anisotropic model and, with no loss of generality, choose: $\alpha_1^2>\alpha_2^2>\alpha_3^2>0$.

The static homogeneous configurations for which the action is extremal are the critical points of $V$ complying
with the constraint:
\begin{equation}
\frac{\partial V}{\partial\phi_a}+2\lambda\phi_a=0 \quad ,\label{eqV}
\end{equation}
where $\lambda$ is the Lagrange multiplier forcing the constraint. There is an important parameter in the system:
\[
\delta^2=\frac{1}{R^2} \left( \frac{1-\alpha_3^2R^2}{\alpha_1^2-\alpha_3^2}+
\frac{\eta_3^2-\eta_1^2}{2(\alpha_1^2-\alpha_3^2)^2}\right) \quad .
\]
If $\delta^2\in (0,1)$, equation (\ref{eqV}) is solved by 18 critical points,
but the following four points:
\begin{equation}
\bar{\phi}_1^2=R^2\delta^2 \quad , \quad \bar{\phi}_2^2\,=\, 0\quad ,\quad \bar{\phi}_3^2=R^2(1-\delta^2)=R^2\bar{\delta}^2\label{delta}
\end{equation}
are the absolute minima of $V$ in ${\mathbb S}^2$. The rest of the critical points are maxima or saddle points. In fact, we will choose $\delta$ in the $0<\delta < 1$ range, because the number of minima is maximized and this circumstance provides a richer structure for the domain wall space. The Lagrange multiplier at the four minima (\ref{delta}) can be easily computed:
\[
\lambda\, =\, \frac{\alpha_3^2 \eta_1^2-\alpha_1^2\eta_3^2}{2(\alpha_1^2-\alpha_3^2)} \quad .
\]

\subsection{Solving the constraint: particle masses}
In order to show explicitly the physical content of the model -symmetry breaking pattern, particle masses,
interaction terms (with and without derivatives) as well as the physical characteristics of the parameters-
it is convenient to solve the constraint by choosing, e.g., $\phi_1$ and $\phi_2$ as independent fields:
$\phi_3^2=R^2-\phi_1^2-\phi_2^2$. On ${\mathbb S}^2$, the unconstrained action becomes:

\begin{eqnarray}
&&S_{{\mathbb S}^2}[\phi_1,\phi_2] = \frac{1}{(\alpha_1^2-\alpha_3^2)^2}\int dx^3\, dt\, \left\{
D_{{\mathbb S}^2}(\partial_\mu\phi_1,\partial_\mu\phi_2, \phi_1, \phi_2)-V_{{\mathbb S}^2}(\phi_1,\phi_2)\right\} \nonumber\\
&&D_{{\mathbb S}^2}(\partial_\mu\phi,\phi)=\frac{1}{2} \left(\partial_\mu\phi_1\partial^\mu
\phi_1+\partial_\mu\phi_2\partial^\mu \phi_2 +  \frac{(\phi_1\partial_\mu
 \phi_1+\phi_2\partial_\mu \phi_2)(\phi_1\partial^\mu \phi_1+\phi_2\partial^\mu \phi_2)}
 {R^2-\phi_1^2-\phi_2^2}\right)\nonumber\\ &&V_{{\mathbb S}^2}(\phi_1,\phi_2)=\frac{1}{2} \left(
\phi_1^2+\sigma^2 \phi_2^2\, -\, \alpha\right)^2 +\frac{\gamma}{2}
\, \phi_1^2+\frac{\beta}{2} \, \phi_2^2 \label{potential4}
\end{eqnarray}
Again, here we have dropped an irrelevant constant and redefined: $x^\mu\to\frac{1}{\alpha_1^2-\alpha_3^2}x^\mu$.
The new parameters are defined in terms of the old ones:
\[
0<\sigma^2=\frac{\alpha_2^2-\alpha_3^2}{\alpha_1^2-\alpha_3^2}<1\,
,\quad \alpha=  \frac{1-\alpha_3^2 \, R^2}{\alpha_1^2-\alpha_3^2}\,
, \quad \gamma=
\frac{\eta_1^2-\eta_3^2}{(\alpha_1^2-\alpha_3^2)^2}\, ,\quad \beta=
\frac{\eta_2^2-\eta_3^2}{(\alpha_1^2-\alpha_3^2)^2}\   .
\]
The range of $\sigma$ is due to the inequalities between the $\alpha_a$'s.

We also restrict the number of independent parameters, for reasons to be explained later, and set{\footnote{Alternatively: $(\alpha_1^2-\alpha_3^2)(\eta_2^2-\eta_3^2)=R^2(\alpha_1^2-\alpha_2^2)+\frac{\alpha_2^2-\alpha_3^2}{\alpha_1^2-\alpha_3^2}
(\eta_1^2-\eta_3^2)$ .}}:
$\beta=\sigma^2(\gamma+\bar{\sigma}^2R^2)$. In a last move, we reshuffle
the forgotten additive constants in such a way that the function $V_{{\mathbb S}^2}$ takes the value of zero at its minima. The potential depends only on $\sigma$, $\delta$
and $R$  ( we shall denote always $\bar{\sigma}^2=1-\sigma^2$ )

\smallskip
\smallskip

\FIGURE[htb]{\includegraphics[height=5cm]{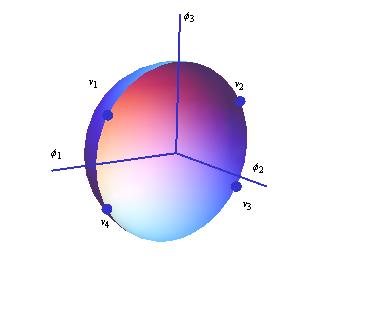}
\caption{Ground states plotted for $\delta=\frac{\sqrt{3}}{2}$, and $R=1$. These values for $R$ and $\delta$ will be maintained in all figures throughout the paper.}}

\smallskip

\begin{equation}
V_{{\mathbb S}^2}(\phi_1,\phi_2)= \frac{1}{2} \left(
\phi_1^2+\sigma^2 \phi_2^2\, -\, \delta^2 R^2\right)^2 \, +\,
\frac{1}{2}\,  R^2 \sigma^2\bar{\sigma}^2  \phi_2^2 \quad .
\label{potential2}
\end{equation}
 The set ${\cal M}$ of zeroes of $V_{{\mathbb S}^2}$, see Figure 1,
\[
{\cal M}\, =\,  \left\{ v_1 \equiv
(R\, \delta\, ,\,
0,R \bar{\delta}), v_2 \equiv (- R\, \delta\, , \,
0,R \bar{\delta}),  v_3\equiv  (-R\, \delta\, ,\,
0,-R\bar{\delta}),v_4 \equiv ( R\, \delta\, ,\,
0,-R \bar{\delta})\right\}  \quad ,
\]
encompasses the four ground states of the model (centered at the constant classical solutions) at the four degenerate absolute minima of $V_{{\mathbb S}^2}(\phi_1, \phi_2)$.

\noindent Expanding $V_{{\mathbb S}^2}$ around any of the vacuum points
\begin{eqnarray*}
V_{{\mathbb S}^2}(R\delta+\phi_1,\phi_2)&=&2\delta^2R^2\phi_1^2+\frac{1}{2}R^2\sigma^2\bar{\sigma}^2\phi_2^2
+\\&+&2\delta R \phi_1(\phi_1^2+\sigma^2\phi_2^2)+\frac{1}{2}(\phi_1^2+\sigma^2\phi_2^2)^2
\end{eqnarray*}
we see that: (a) The particle masses are: $\mu_1^2=4\delta^2R^2$, $\mu_2^2=\sigma^2\bar{\sigma}^2R^2$.
(b) The ${\mathbb Z}_2\times {\mathbb Z}_2$ symmetry engendered by $\phi_1\to -\phi_1$ and $\phi_2\to -\phi_2$
is spontaneously broken to the last ${\mathbb Z}_2$ subgroup by the choice of vacuum. (c) There are two trivalent
vertices with couplings $2\delta R$ and $2\delta R \sigma^2$. (d) There are three tetravalent vertices with couplings
$\frac{1}{2}$, $\sigma^2$, and $\frac{\sigma^4}{2}$.

The r$\hat{\rm o}$le of $\frac{1}{R^2}$ as a coupling constant comes from the following expansion
of the $\partial_\mu\phi_3\partial^\mu\phi_3$ term in the Lagrangian:
\begin{eqnarray*}
&&\frac{(\phi_1\partial_\mu\phi_1+\phi_2\partial_\mu\phi_2
)}{\sqrt{R^2-\phi_1^2-\phi_2^2}}\cdot\frac{(\phi_1\partial^\mu\phi_1+\phi_2\partial^\mu\phi_2
)}{\sqrt{R^2-\phi_1^2-\phi_2^2}}\simeq
{1\over R^2}\left( \phi_1\frac{\partial\phi_1}{\partial
x^\mu}+\phi_2\frac{\partial\phi_2}{\partial x^\mu}\right)\left(
\phi_1\frac{\partial\phi_1}{\partial
x_\mu}+\phi_2\frac{\partial\phi_2}{\partial x_\mu}\right)+
\\&&+\frac{1}{R^4}(\phi_1^2+\phi_2^2)\left(
\phi_1\frac{\partial\phi_1}{\partial
x^\mu}+\phi_2\frac{\partial\phi_2}{\partial x^\mu}\right)\left(
\phi_1\frac{\partial\phi_1}{\partial
x_\mu}+\phi_2\frac{\partial\phi_2}{\partial
x_\mu}\right)+ \cdots \qquad ,
\end{eqnarray*}
i.e., an infinite number of vertices with two field derivatives arise proportional to powers of $\frac{1}{R^2}$
due to the geometry of the system.

\subsection{Solitonic domain walls }

The non-linear field equations of the system are:
\begin{equation}
 \partial_0^2 \, \phi_a-\nabla^2
 \phi_a=-2 \alpha_a^2 \phi_a \left( \sum_{b=1}^3 \alpha_b^2
\phi_b^2 - 1 \right)-\eta_a^2
\phi_a\, +\lambda\, \phi_a \, , \quad a=1,2,3.\label{eqq0}
\end{equation}
We temporarily return to keep $\phi_3$ explicit. $\lambda$ is again the Lagrange multiplier in the equation (\ref{eqq0}). For any solution of (\ref{eqq0}) it can be shown to be{\footnote{Multiply (\ref{eqq0}) by $\phi_a$ and sum the three equations. Also use that: $\sum_{a=1}^3\phi_a\partial_\mu\phi_a=0$ to perform a partial integration.}}:
\begin{equation}
\lambda\, =\, \frac{1}{R^2}\,  \, \sum_{a=1}^3 \left( -(\partial_0\phi_a)^2+\vec{\nabla}\phi_a\cdot\vec{\nabla}\phi_a\, +\, \phi_a\cdot \frac{\delta V}{\delta\phi_a}\right) \, \, \, , \quad  \vec{\nabla}=\frac{\partial}{\partial x^1}\vec{i}_1+\frac{\partial}{\partial x^2}\vec{i}_2+\frac{\partial}{\partial x^3}\vec{i}_3\quad ,
\end{equation}
where $\vec{\nabla}$ is the gradient in the spatial subspace of Minkowski space.

Our main goal in this paper is to investigate the domain wall solutions in this model.
{\bf Domain walls} are non-singular solutions of the field equations (\ref{eqq0}) such that their energy density has a space-time dependence of the form: ${\cal E}(x^0,x^1,x^2,x^3)={\cal E}(x^1-vx^0)$, where $v$ is
some velocity vector in the $x^1$ direction, and their energy functional:
\begin{eqnarray*}
E[\vec{\phi}]&=&\lim_{L\to\infty}\frac{L^2}{\alpha_1^2-\alpha_3^2}\displaystyle  \int dx^1 \left( \frac{1}{2} \,
\partial_0\vec{\phi}\cdot \partial_0 \vec{\phi}+
\frac{1}{2}\,  \partial_1\vec{\phi}\cdot \partial_1
\vec{\phi}+V(\phi_1,\phi_2,\phi_3)\right) \, \\ &=& \, \lim_{L\to\infty}\frac{L^2}{\alpha_1^2-\alpha_3^2}\, \int dx^1 \, \,
{\cal E}(x^0,x^1) \quad ,
\end{eqnarray*}
is proportional to the area $L^2$ of a normalizing square in the $x_2-x_3$ plane. Therefore, these solutions will be domain walls or solitonic (thick) 2-branes orthogonal to the $x^1$-axis.
The Lorentz invariance of the model implies that it suffices to know the $x^0$-independent
solutions $\vec{\phi}(x^1)$ in order to obtain
the domain walls of the model:
$\vec{\phi}(x^0,x^1)=\vec{\phi}(x^1-vx^0)$. For static and $x^2$-, $x^3$- independent
configurations the PDE system (\ref{eqq0}) becomes the following system of three
ordinary differential equations:
\begin{equation}
\frac{d^2 \phi_a}{d(x^1)^2}\, =\, -2 \alpha_a^2 \phi_a \left( \sum_{b=1}^3 \alpha_b^2
\phi_b^2 - 1 \right)^2-\eta_a^2
\phi_a\, +\lambda\, \phi_a\, ,\quad a=1,2,3 \label{eqqq}
\end{equation}
and the tension of the wall (2-brane) reduces to:
\[
\Omega(\vec{\phi})=\lim_{L\to\infty}\frac{E[\vec{\phi}]}{L^2}=\frac{1}{\alpha_1^2-\alpha_3^2}\int dx^1 \left( \frac{1}{2}\,  \frac{d
\vec{\phi}}{dx^1}\cdot \frac{d\vec{\phi}}{dx^1}\, +\,
V(\vec{\phi})\right)\, =\, \frac{1}{\alpha_1^2-\alpha_3^2}\int dx^1 \, \, {\cal E}(x^1) \quad .
\]

The ODE system (\ref{eqqq}) can be interpreted as the Newton equations of a mechanical system, which we shall refer to as the analogous mechanical system to our field theoretical problem. Thus, the $x^1$ coordinate in ${\mathbb R}^3$ will be identified with $\tau$: the mechanical time. The field configurations $\phi_a(x^1)$ will give the paths in ${\mathbb S}^2$, $X_a(\tau)$. The $V(\phi_1,\phi_2,\phi_3)$ function of the field theory will be minus the mechanical potential $V(X_1,X_2,X_3)$. Finally, the domain wall tension $\Omega$ will be interpreted as the mechanical action functional. We shall always use, however, the field theoretical notation, although the interpretation should be clear.
It should be stressed that the mechanical potential is minus the function $V$.

The finite wall tension (finite mechanical action) requirement is fulfilled if and only if the asymptotic conditions
hold:
\begin{equation}
\lim_{x^1\to \pm \infty} \, \frac{d \vec{\phi}}{dx^1}\, =\, 0\quad
,\qquad \lim_{x^1\to \pm \infty} \, \vec{\phi}\, \in {\cal M} \,  .\label{asy}
\end{equation}
Thus, the space of finite wall tension configurations
\[
{\cal C}=\left\{ \phi \in {\rm Maps}({\mathbb R}\times{\mathbb R^2},{\mathbb S}^{2})/{\rm Maps}({\mathbb R^2},{\rm point}): \Omega[\phi]<+\infty \right\}\,=\,  \bigcup_{i,j=1}^4 {\cal C}_{ij}
\]
is the union of sixteen disconnected sectors: $ {\cal C}_{ij}$, labeled by the element
of ${\cal M}$ reached by each configuration at $x^1\to -\infty$ and
$x^1\to \infty$. If $i\neq j$, the finite tension walls will be termed as topological walls, whereas
non-topological walls will be the solutions belonging to the ${\cal C}_{ii}$ sectors.

\subsection{Trial orbits and two basic walls}

Before searching for general domain wall solutions
of the ODE system (\ref{eqqq}), we shall show two particular ones by sticking to Rajaraman's trial orbit method \cite{Rajaraman}.

\FIGURE{\includegraphics[height=4.5cm]{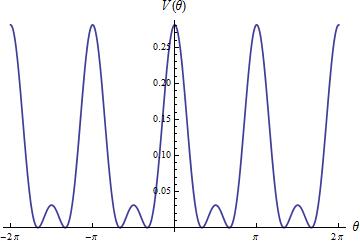}\caption{Graphics of $V_{{\mathbb S}^1}(\theta)$.}}

In the meridian $\phi_2=0$, which we choose as a trial orbit, the constraint becomes
$\phi_1^2+\phi_3^2\, =\, R^2$. The polar angle in this maximal circle solves the constraint
\[
\phi_1(x)=R \, \sin \theta(x)\quad,\quad \phi_3(x)=R\cos\theta(x) \qquad .
\]
We have written $x\equiv x^1$ for simplicity and will maintain this convention in the rest of the paper.
The ODE system (\ref{eqqq}) reduces to the single second-order OD equation:
\begin{equation}
\frac{d^2\theta(x)}{d x^2}\, =\, R^2 \, \sin 2\theta(x)\, \left( \sin^2\theta(x)-\delta^2\right)\label{eqtheta}\qquad .
\end{equation}

\noindent The mechanical potential on the orbit is:
\[
U(\theta)=-V_{{\mathbb S}^1}(\theta)=-\frac{R^4}{2}\left({\rm sin}^2\theta(x)-\delta^2\right)^2 \quad .
\]
\noindent The mechanical energy $I$ provides a first-integral for (\ref{eqtheta}):
\[
I=\frac{1}{2}\left(\frac{d\theta}{dx}\right)^2+\frac{1}{R^2}U(\theta)=\frac{1}{2}\left(\frac{d\theta}{dx}\right)^2-\frac{R^2}{2}\left({\rm sin}^2\theta(x)-\delta^2\right)^2 \label{eqtheta1} \qquad .
\]
 The critical points of $U$ are: (1) minima: $\theta_0=0$, the North pole, and $\theta_\pi=\pi$, the South pole and $\theta_\frac{\pi}{2}=\frac{\pi}{2}$ and $\theta_\frac{3\pi}{2}=\frac{3\pi}{2}$ antipodal points in the equator. (2) maxima: $\theta_{+0}={\rm arcsin}\delta$, $\theta_{-0}=-{\rm arcsin}\delta$,
$\theta_{+\pi}={\rm arcsin}\delta+\pi$, $\theta_{-\pi}=-{\rm arcsin}\delta+\pi$, respectively, the $v_1$, $v_2$, $v_3$, and $v_4$ minima of the field theory. The mechanical energy for all these maxima is $I=0$, which must therefore be
the value of the integration constant of (\ref{eqtheta}) required to obtain solutions with finite wall tension.
Therefore, the topological wall solutions correspond to the quadratures of
\[
\frac{d\theta}{dx}=\pm R\left({\rm sin}^2\theta-\delta^2\right) \qquad ,
\]
which produce two types of analytical outcomes:

\medskip

\noindent {\bf 1.- Polar Meridian Domain Walls (PMW):} We shall denote this kind of solutions as $\theta_{12}^{\rm PMW}(x)$,  $\theta_{21}^{\rm PMW}(x)$, $\theta_{34}^{\rm PMW}(x)$ and $\theta_{43}^{\rm PMW}(x)$, where the indexes stand for the asymptotically connected vacua via the domain wall. For example, the orbit of the solution $\theta_{12}^{\rm PMW}$ corresponds to the piece of the $\phi_2=0$ meridian joining the $v^1$ and $v^2$ vacua, which crosses the North Pole. $\theta_{21}^{\rm PMW}$ is the anti-wall of the previous solution while $\theta_{34}^{\rm PMW}$ and $\theta_{43}^{\rm PMW}$ are similar domain walls confined to the South Polar Region. We find that
\[
\theta_{12}^{\rm PMW}(x) = \arcsin \left( \frac{\delta \sinh [R\delta \bar{\delta} (x-x_0)]}{\sqrt{\cosh^2[R\delta\bar{\delta} (x-x_0)]-\delta^2}}\right) \, \, \, , \quad \theta_{12}^{\rm PMW}(x)\in (\theta_{-0},\theta_{+0}) \, \, ,
\]
and therefore $\theta_{21}^{\rm PMW}(x)=\theta_{12}^{\rm PMW}(-x)$, $\theta_{34}^{\rm PMW}(x)=\theta_{12}^{\rm PMW}+\pi\in (\theta_{-\pi},\theta_{+\pi})$, and $\theta_{43}^{\rm PMW}(x)=\theta_{34}^{\rm PMW}(-x)$. These topological walls live in the topological sectors ${\cal C}_{12}$, ${\cal C}_{34}$, ${\cal C}_{21}$, and ${\cal C}_{43}$. All of them have the same tension, $\Omega({\rm PMW}) = \Omega(\theta_{12})=\Omega(\theta_{21}) = \Omega(\theta_{34})=\Omega(\theta_{43})$, where
\[
\Omega({\rm PMW})= \frac{R^3}{\alpha_1^2-\alpha_3^2}\left[\delta\bar\delta-(1-2\delta^2)
\arccos \bar\delta \right]\quad .
\]
We write all these topological walls in a unified way in the original field variables
\begin{eqnarray}
\phi_1^{\rm PMW}(x;\epsilon_1)&=&  \frac{(-1)^{\epsilon_1} R\delta \sinh \left[ R \delta \bar{\delta} (x-x_0)\right]}{\sqrt{\cosh^2 \left[R \delta \bar{\delta} (x-x_0)\right] -\delta^2}}\, \, \, ,\qquad    \phi_2^{\rm PMW}(x)=0\nonumber \\   \phi_3^{\rm PMW}(x;\epsilon_3)&=& \frac{(-1)^{\epsilon_3}R\bar{\delta}  \cosh \left[ R \delta \bar{\delta} (x-x_0)\right]}{\sqrt{\cosh^2 \left[ R \delta\bar{\delta} (x-x_0)\right] -\delta^2}} \, \, \, , \qquad {\epsilon_1} , {\epsilon_3}=0,1 \quad .\label{kinks1}
\end{eqnarray}
For ${\epsilon_3}=0$ the walls belong to ${\cal C}_{12}$ (${\epsilon_1}=1$) and ${\cal C}_{21}$ (${\epsilon_1}=0$).
For ${\epsilon_3}=1$ the walls belong to ${\cal C}_{34}$ (${\epsilon_1}=1$) and ${\cal C}_{43}$ (${\epsilon_1}=0$).
The $\phi_1$-component has the form of a kink and the $\phi_3$-component is bell shaped.
\FIGURE{\includegraphics[height=3.5cm]{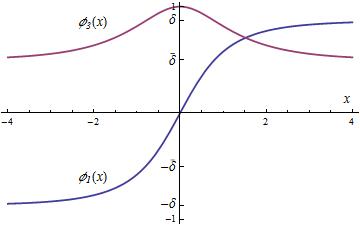}\qquad \includegraphics[height=4cm]{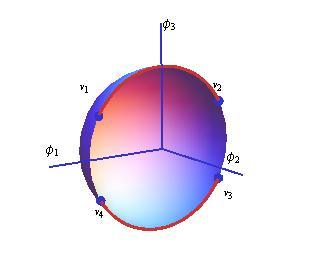}
\caption{Field profiles (\ref{kinks1}) for ${\epsilon_1}=0={\epsilon_3}$, $x_0=0$ (left). Orbits (red curves) in ${\cal C}_{21}$, ${\cal C}_{12}$ and ${\cal C}_{43}$, ${\cal C}_{34}$ (right).
}}

\noindent {\bf 2.- Tropical Meridian Domain Walls (TMW):} We shall now denote these solutions as $\theta_{14}^{\rm TMW}(x)$, $\theta_{41}^{\rm TMW}(x)$, $\theta_{23}^{\rm TMW}(x)$ and $\theta_{32}^{\rm TMW}(x)$. For example, $\theta_{14}^{\rm TMW}(x)$ connects from the vacuum $v^1$ to the $v^4$ vacuum crossing the equator of the sphere. We find that
\[
\theta_{41}^{\rm TMW}(x) = \arccos \left( \frac{\bar{\delta} \sinh [R\delta \bar{\delta} (x-x_0)]}{\sqrt{\cosh^2[R\delta\bar{\delta} (x-x_0)]-\bar{\delta}^2}}\right)\, \, \, , \quad \theta_{41}^{\rm TMW}(x)\in (\theta_{-\pi},\theta_{+0})
\]
and therefore $\theta_{14}^{\rm TMW}(x)=\theta_{41}^{\rm TMW}(-x)$, $\theta_{32}^{\rm TMW}(x)=\theta_{41}^{\rm TMW}(x)+\pi \in (\theta_{+\pi},\theta_{-0})$ and $\theta_{23}^{\rm TMW}(x)=\theta_{32}^{\rm TMW}(-x)$. These topological walls belong to the ${\cal C}_{41}$, ${\cal C}_{32}$, ${\cal C}_{14}$, and ${\cal C}_{23}$
sectors and their tension is $\Omega({\rm TMW})=\Omega(\theta_{14}^{\rm TMW})=\Omega(\theta_{41}^{\rm TMW})=\Omega(\theta_{23}^{\rm TMW})=\Omega(\theta_{32}^{\rm TMW})$ where
\[
\Omega({\rm TMW})=\frac{R^3}{\alpha_1^2-\alpha_2^2}\left[\delta\bar\delta+(1-2\delta^2)
\arccos \delta \right]\quad .
\]
In the original field coordinates the analytical expressions are:
\begin{eqnarray}
\phi_1^{\rm TMW}(x;\kappa_1)&=& \frac{(-1)^{\kappa_1} R\delta \cosh \left[ R \delta \bar{\delta} (x-x_0)\right]}{\sqrt{\cosh^2 \left[R \delta \bar{\delta} (x-x_0)\right] -\bar{\delta}^2}}\, \, \, , \qquad   \phi_2^{\rm TMW}(x)=0\nonumber  \\  \phi_3^{\rm TMW}(x;\kappa_3)&=& \frac{(-1)^{\kappa_3} R\bar{\delta} \sinh \left[ R \delta \bar{\delta} (x-x_0)\right]}{\sqrt{\cosh^2 \left[ R \delta\bar{\delta} (x-x_0)\right] -\bar{\delta}^2}} \, \, \, , \qquad \kappa_1, \kappa_3=0,1 \quad .\label{kinks2}
\end{eqnarray}
If $\kappa_3=0$, the domain walls live in ${\cal C}_{41}$ ($\kappa_1=0$) and ${\cal C}_{14}$ ($\kappa_1=1$). If
$\kappa_3=1$ the domain walls live in ${\cal C}_{32}$ ($\kappa_1=0$) and ${\cal C}_{23}$ ($\kappa_1=1$). The $\phi_1$
component is bell shaped and the $\phi_3$ now has the form of a kink.

\FIGURE{\includegraphics[height=3.5cm]{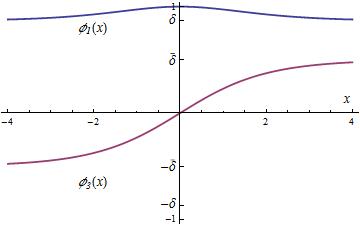}\  \qquad    \includegraphics[height=4cm]{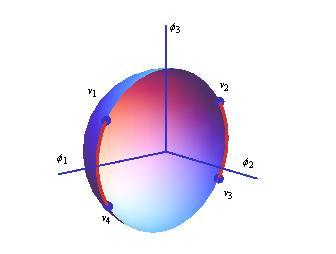}
\caption{Field profiles (\ref{kinks2}) for $\kappa_1=0=\kappa_3$, $x_0=0$ (left). Orbits (red curves) in ${\cal C}_{41}$, ${\cal C}_{14}$ and ${\cal C}_{23}$, ${\cal C}_{32}$ (right).}}

To end this subsection we show ( Figure 5 ) the tension densities of these two kinds of topological wall.

\FIGURE{\includegraphics[height=4cm]{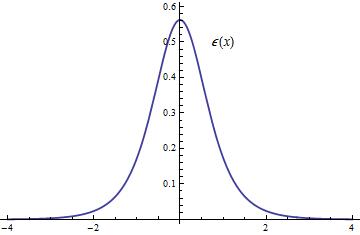}\ \quad \includegraphics[height=4cm]{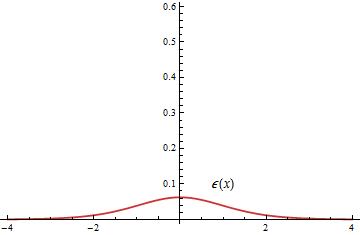}
\caption{Cross-sections of energy densities: (left) PMW walls in ${\cal C}_{12}$ and ${\cal C}_{21}$. (right) TMW walls in ${\cal C}_{23}$ and ${\cal C}_{14}$. }}

\section{Hamilton-Jacobi separability}
\subsection{Sphero-conical coordinates}
In order to search for all the domain wall solutions of the model, we introduce sphero-conical coordinates $(\lambda_0,\lambda_1,\lambda_2)$ in ${\mathbb R}^3$, see, e.g., \cite{Neumann}-\cite{Moser}:
\[
\phi_1^2= \lambda_0 \frac{(a_1-\lambda_1)(a_1-\lambda_2)
}{(a_1-a_2) (a_1-a_3)}\, ,\quad  \phi_2^2= \lambda_0
\frac{(a_2-\lambda_1)(a_2-\lambda_2)}{(a_2-a_1)
(a_2-a_3)}\, ,\quad  \phi_3^2= \lambda_0
\frac{(a_3-\lambda_1)(a_3-\lambda_2) }{(a_3-a_1) (a_3-a_2)}
\]
with separation constants related to the $\sigma$-parameter:
\[
a_1=0\, ,\  a_2=\bar{\sigma}^2\, ,\  a_3=1 \quad \Leftrightarrow \quad 0<\lambda_1<\bar{\sigma}^2 <\lambda_2<1
\]
In this system of coordinates, the constraint is simply $
\lambda_0=\phi_1^2+\phi_2^2+\phi_3^2=R^2$,
such that the field components restricted to ${\mathbb S}^2$ in terms of sphero-conical coordinates read:
\begin{equation}
\phi_1^2= \frac{R^2}{\bar{\sigma}^2} \, \lambda_1\, \lambda_2\, \, \, , \quad
\phi_2^2= \frac{R^2}{\sigma^2 \bar{\sigma}^2} \, (\bar{\sigma}^2-\lambda_1)
(\lambda_2-\bar{\sigma}^2)\, \, \, , \quad \phi_3^2= \frac{R^2}{\sigma^2} \,
(1-\lambda_1)(1-\lambda_2)\quad .\label{spcf}
\end{equation}
The map induced by the change of coordinates (\ref{spcf}) is eight-to-one; i.e., each octant of the ${\mathbb S}^2$ sphere is mapped onto the rectangle $P_2$ in the $(\lambda_1,\lambda_2)$-plane. See Figure 6 for $\sigma=\frac{1}{\sqrt{2}}$, a selection of $\sigma$ maintained in all the graphics below.

\FIGURE{\includegraphics[height=4cm]{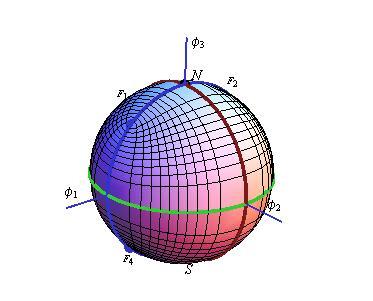}    \qquad       \includegraphics[height=4cm]{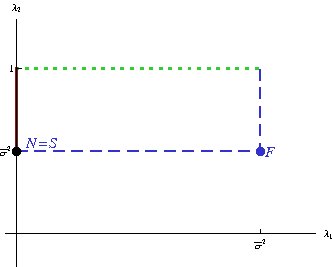}
\caption{The ${\mathbb S}^2$ sphere in ${\mathbb R}^3$ (left). The $P_2$
rectangle in the $(\lambda_1,\lambda_2)$-plane (right). The dotted green line of $P_2$ is mapped to the equator while the dashed blue line of $P_2$ is mapped to the meridian which crosses the foci. The red solid line in $P_2$ corresponds to the other meridian displayed in the sphere. The $\lambda_1=$ constant and $\lambda_2=$ constant iso-curves are shown back in ${\mathbb S}^2$.}}

The sphero-conical coordinates distinguish four special points in ${\mathbb S}^2$: the foci $F_1$, $F_2$, $F_3$ and $F_4$, all of them mapped onto the corner $(\lambda_1,\lambda_2)\, =\, (\bar{\sigma}^2,\bar{\sigma}^2)$ of $P_2$. There is a direct relation between these coordinates and the elliptical coordinates on a sphere used in \cite{AMAJ}-\cite{AMAJ1}. Choosing two non-antipodal foci, for instance $F_1$ and $F_2$, we have that:
\[
\lambda_1=\sin^2 \left(\frac{r_1-r_2}{2R}\right)\quad,\qquad \lambda_2=\sin^2 \left(\frac{r_1+r_2}{2R}\right)
\]
\FIGURE{\includegraphics[height=4cm]{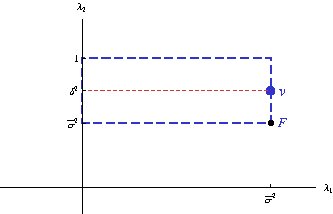}
\caption{The ground states and the foci in sphero-conical coordinates.}}
\smallskip

\noindent where $r_1$ and $r_2$ are the geodesic distances from a given point in ${\mathbb S}^2$ to $F_1$ and $F_2$ respectively. $v=\frac{r_1-r_2}{2}$ and $u=\frac{r_1+r_2}{2}$ are the spherical-elliptic coordinates in ${\mathbb S}^2$. Thus the iso-curves depicted in Figure 6 (left) represent ``ellipses" and ``hyperbolas" on ${\mathbb S}^2$.

The eight-to-one correspondence between ${\mathbb S}^2$ and $P_2$ maps all the ${\cal C}_{ij}$ sectors onto only one; also, the four points $v_1, v_2,v_3$ and $v_4$ are mapped onto the point $v\equiv (\bar{\sigma}^2,\delta^2)$ in $P_2$. See Figure 7.

We are assuming that $\bar{\sigma}^2< \delta^2$. The choice $\bar{\sigma}^2> \delta^2$ is equivalent to this modulo a
$\frac{\pi}{2}$ rotation around the $\phi_2$-axis.

It is easy to check that the non-derivative part of the field theoretical energy density, the $V_{{\mathbb S}^2}$-function (\ref{potential4}), in sphero-conical coordinates reads:
\[
V_{{\mathbb S}^2}(\lambda_1,\lambda_2)=
\frac{R^4}{2(\lambda_2-\lambda_1)} \left(
(\bar{\sigma}^2-\lambda_1)
(\delta^2-\lambda_1)^2+(\lambda_2-\bar{\sigma}^2)
(\lambda_2-\delta^2)^2\right) \label{HJpot}
\]
if and only if $\beta=\sigma^2\, (\gamma+\bar{\sigma}^2 R^2)$.

The action in sphero-conical coordinates is
\[
S=\int \, d^4x \, {\cal L}=\int \, d^4x \, \left\{ \frac{1}{2}g_{11}(\lambda_1,\lambda_2)\partial_\mu\lambda_1\partial^\mu\lambda_1 +\frac{1}{2}g_{22}(\lambda_1,\lambda_2)\partial_\mu\lambda_2\partial^\mu\lambda_2- V_{{\mathbb S}^2}(\lambda_1,\lambda_2)\right\}\qquad ,
\]
where the components of the metric tensor induced in $P_2$ by the change of coordinates are:
\[
g^{11}(\lambda_1,\lambda_2)=g_{11}^{-1} = \frac{ 4 \lambda_1 (\bar{\sigma}^2
-\lambda_1)(1-\lambda_1)}{R^2(\lambda_2-\lambda_1)} \  ,
\quad g^{22}(\lambda_1,\lambda_2)=g_{22}^{-1}= \frac{ 4 \lambda_2 (\lambda_2-\bar{\sigma}^2)(1-\lambda_2)}{R^2(\lambda_2-\lambda_1)} \quad .
\]
Defining
\[
\pi^\mu_1=\frac{\delta{\cal L}}{\delta\partial_\mu\lambda_1}=g_{11}(\lambda_1,\lambda_2)\partial^\mu\lambda_1 \, \, , \quad \pi^\mu_2=\frac{\delta{\cal L}}{\delta\partial_\mu\lambda_2}=g_{22}(\lambda_1,\lambda_2)\partial^\mu\lambda_2 \quad ,
\]
one sees that the energy-momentum tensor $T^{\mu\nu}=\pi^\mu_1\partial^\nu\lambda_1+\pi^\mu_2\partial^\nu\lambda_2-
g^{\mu\nu}{\cal L}$ of the wall solutions of (\ref{eqqq}) is diagonal, in the form:
\[
T_\mu^\nu (x)=\left(g_{11}\frac{\partial\lambda_1}{\partial x}\frac{\partial\lambda_1}{\partial x}+g_{22}\frac{\partial\lambda_2}{\partial x}\frac{\partial\lambda_2}{\partial x}\right){\rm diag}(1,0,1,1)\, \, \, .
\]
$T^1_1(x)=0$ is due to the continuity equation $\partial_x T^1_1(x)=0$ and the remaining components are of the form
$T^\nu_\mu(x)\propto \delta^\nu_\mu$ because of the parallel unbroken Lorentz invariance (in the $x^2$-,
$x^3$-directions) of the wall.
\subsection{The analogous mechanical system}
The mechanical momenta are: $\pi_1=g_{11}(\lambda_1,\lambda_2)\frac{d \lambda_1}{d x}$, $\pi_2=g_{11}(\lambda_1,\lambda_2)\frac{d \lambda_2}{d x}$  . The mechanical Hamiltonian
\[
{\cal H}=\frac{2\lambda_1(\bar{\sigma}^2-\lambda_1)(1-\lambda_1)}{R^2(\lambda_2-\lambda_1)}\pi_1^2 +
\frac{2\lambda_2(\lambda_2-\bar{\sigma}^2)(1-\lambda_2)}{R^2(\lambda_2-\lambda_1)}\pi_2^2 +U(\lambda_1,\lambda_2)\quad ,
\]
$U(\lambda_1,\lambda_2)=-V_{{\mathbb S}^2} (\lambda_1,\lambda_2)$, is of the St$\ddot{\rm a}$ckel form and the mechanical system is Hamilton-Jacobi separable in the variables $\lambda_1$, $\lambda_2$.
This is the reason of our choice of $\beta$: to cope with a new, previously unknown, integrable mechanical system belonging to the class of the Neumann problem \cite{Neumann}.

The solutions of this mechanical system complying with the asymptotic conditions
\begin{equation}
\lim_{x\to \pm \infty} \, \frac{d \lambda_1(x)}{dx}\, =\, \lim_{x\to \pm \infty} \, \frac{d \lambda_2(x)}{dx}=0 \quad
,\qquad \lim_{x\to \pm \infty} \, (\lambda_1(x),\lambda_2(x))\, =\, (\bar{\sigma}^2,\delta^2) \,  .\label{asy2}
\end{equation}
forced by (\ref{asy}), will provide all the domain wall solutions with finite tension of the field theoretical model.

In sphero-conical coordinates the wall tension is:
\[
\Omega(\lambda_1,\lambda_2) = \frac{1}{\alpha_1^2-\alpha_3^2} \int d x \left[
\frac{1}{2} g_{11}(\lambda_1,\lambda_2) \left( \frac{ d \lambda_1}{ d x} \right)^2 +
\frac{1}{2} g_{22}(\lambda_1,\lambda_2)\left( \frac{ d \lambda_2}{ d x} \right)^2 + V_{{\mathbb S}^2}(\lambda_1,\lambda_2)\right]
 \]
If a solution, $W(\lambda_1, \lambda_2)$, of the PDE:
\begin{equation}
V_{{\mathbb S}^2}(\lambda_1,\lambda_2)= \frac{1}{2} \left( g^{11} \left(
\frac{ \partial W}{\partial \lambda_1}\right)^2 +
 g^{22} \left(
\frac{ \partial W}{\partial \lambda_2}\right)^2 \right)\label{HJPDE}
\end{equation}
is known, the Bogomolnyi arrangement \cite{Bogomolny} of the wall tension:
\[
\Omega(\lambda_1,\lambda_2)= \frac{1}{\alpha_1^2-\alpha_3^2} \int d x  \frac{1}{2}
\sum_{i=1}^2 g_{ii} \left( \frac{ d \lambda_i}{d x} - g^{ii}
\frac{\partial W}{\partial \lambda_i} \right)^2 + \frac{1}{\alpha_1^2-\alpha_3^2}
\int d x \sum_{i=1}^2 \frac{\partial W}{\partial\lambda_i} \frac{ d
\lambda_i}{d x}
\]
shows that the absolute minima of the wall tension functional are the solutions of the first-order ODE system
\begin{equation}
\frac{ d \lambda_1}{d x } =
g^{11}(\lambda_1,\lambda_2) \frac{\partial W}{\partial \lambda_1}\, \, \, , \quad
\frac{ d \lambda_2}{d x } = g^{22}(\lambda_1,\lambda_2) \frac{\partial W}{\partial
\lambda_2} \label{difsup} \quad ,
\end{equation}
complying with the asymptotic conditions (\ref{asy2}).

 However, the PDE (\ref{HJPDE}) is no more than the time-independent Hamilton-Jacobi equation of the analogous mechanical problem (for zero {\it mechanical energy}), such that $W(\lambda_1,\lambda_2)$ is the Hamilton's characteristic function. Separability implies the existence of solutions of the form:
\[
W(\lambda_1,\lambda_2)=W_1(\lambda_1)+W_2(\lambda_2)
\]
Integration of the Hamilton-Jacobi equation in general involves hyper-elliptical integrals, but the quadratures  reduce to simple irrational integrals for the solutions of finite (mechanical) action in infinite (mechanical) time that
provide finite tension domain walls.

The mechanical system is completely integrable in the Arnold-Liouville sense, admitting two integrals of motion in involution:
\begin{eqnarray*}
I_1&=&\frac{1}{2} g^{11} \, \pi_1^2+\frac{1}{2} g^{22} \, \pi_2^2 -\frac{R^4}{2(\lambda_2-\lambda_1) }\left( (\bar{\sigma}^2-\lambda_1)(\delta^2-\lambda_1)^2+(\lambda_2-\bar{\sigma}^2)(\lambda_2-\delta^2)^2\right)\\
I_2&=& \frac{1}{2} g^{11} \lambda_2 \, \pi_1^2+\frac{1}{2} \, g^{22}\, \lambda_1 \, \pi_2^2 -\frac{R^4}{2} \left(  \lambda_1 \lambda_2 \left( \lambda_1+\lambda_2-\bar{\sigma}^2-2\delta^2\right)-\bar{\sigma}^2 \delta^4\right)\quad .
\end{eqnarray*}
The finite tension conditions (\ref{asy2}) require that $I_1=I_2=0$ and the first-order PDE (\ref{HJPDE}) becomes the ODE system:
\begin{equation}
\frac{ d W_1}{d \lambda_1} = (-1)^{\epsilon_1} \frac{ R^{3}}{2}
\frac{ (\delta^2-\lambda_1)}{\sqrt{\lambda_1(1-\lambda_1)}} \, ,\quad
\frac{ d W_2}{d \lambda_2} = (-1)^{\epsilon_2} \frac{ R^{3}}{2}
\frac{ (\delta^2-\lambda_2)}{\sqrt{\lambda_2(1-\lambda_2)}} \quad . \label{rfods}
\end{equation}
The solution of (\ref{rfods})
\begin{eqnarray*} W(\lambda_1,\lambda_2) &=& \frac{(-1)^{\epsilon_1}
R^{3}}{2} \left(  \sqrt{\lambda_1(1-\lambda_1)} +
(1 -2\delta^2) \arctan \sqrt{1-\lambda_1 \over \lambda_1}
\right) \nonumber \\ &+& \frac{(-1)^{\epsilon_2} R^{3}}{2} \left(
\sqrt{\lambda_2(1-\lambda_2)} + (1 -2\delta^2) \arctan
\sqrt{1-\lambda_2 \over \lambda_2} \right) \, \, , \quad {\epsilon_1},{\epsilon_2} = 0,1\label{superpot1}
\end{eqnarray*}
is the complete integral of the HJ equation (\ref{HJPDE}) such that the ODE system (\ref{difsup}) reads
\begin{eqnarray}
\frac{ d \lambda_1}{d x} &=& - R (-1)^{\epsilon_1}
\frac{ 2 (\bar{\sigma}^2 -\lambda_1) (\delta^2-\lambda_1) \sqrt{\lambda_1(1-\lambda_1)}}{\lambda_1-\lambda_2} \label{eq1}\\
\frac{ d \lambda_2}{d x} &=& - R (-1)^{\epsilon_2} \frac{ 2
(\bar{\sigma}^2 -\lambda_2) (\delta^2-\lambda_2)
\sqrt{\lambda_2(1-\lambda_2)}}{\lambda_2-\lambda_1} \label{eq2} \quad ,
\end{eqnarray}
encoding all the domain wall solutions as the separatrix trajectories between bounded and unbounded
motions in the analogous mechanical system. Note that the flow induced by the gradient of $W$ is:
\[
\frac{d\lambda_2}{d\lambda_1}=(-1)^{{\epsilon_1}-{\epsilon_2}}
\frac{(\bar\sigma^2-\lambda_2)(\delta^2-\lambda_2)\sqrt{\lambda_2(1-\lambda_2)}}
{(\bar\sigma^2-\lambda_1)(\delta^2-\lambda_1)\sqrt{\lambda_1(1-\lambda_1)}} \qquad ;
\]
i.e., the flow is only undefined at the vacuum $(\lambda_1=\bar\sigma^2, \lambda_2=\delta^2)$, and
the focus $(\lambda_1=\bar\sigma^2, \lambda_2=\bar\sigma^2)$, points where infinite orbits meet or cross.

To end this subsection we remark that we must bear in mind two different zones in the $(\lambda_1,\lambda_2)$-plane in order to search for the most general wall solutions. The two zones are delimited by the straight line $\lambda_2=\delta^2$. In the original variables this line is given by the intersection between the conical surface $\frac{\phi_1^2}{\delta^2}+ \frac{\phi_2^2}{\delta^2-\bar{\sigma}^2} - \frac{\phi_3^2}{1-\delta^2}=0$ and the sphere $\phi_1^2+\phi_2^2+\phi_3^2=R^2$, which determines two ellipses described on each of the hemispheres. We shall refer to these curves as \textit{tropical ellipses}, using a geographical analogy in the sphero-conical coordinates. These so-called tropical ellipses divide the sphere into three regions, the intertropical zone (characterized by $\lambda_2 >\delta^2$ in the sphero-conical plane) and the North Polar zone, or ``arctic" region, and the South Polar zone, or ``ant-arctic" region, (characterized by $\lambda_2 < \delta^2$ in the sphero-conical plane).

\subsection{One more basic wall: Trial orbit in $P_2$}

The orbit $\lambda_2 =\delta^2$, see Figure 8 (left), solves
the equation (\ref{eq2}) and the equation (\ref{eq1}) on this orbit can readily be integrated. We find domain walls that live on the tropical ellipses
\begin{equation}
\lambda_1^{\rm TW}(x) = \frac{\bar{\sigma}^2\sinh^2 \left[ R \sigma\bar{\sigma} (x-x_0)\right]}{\cosh^2 \left[ R  \sigma\bar{\sigma} (x-x_0)\right]-\bar{\sigma}^2} \quad\quad \lambda_2^{\rm TW} =\delta^2 \quad , \label{tk3sc}
\end{equation}
which we will refer to as {\bf Tropical Domain Walls} (TW). It should be understood that (\ref{tk3sc}) solves (\ref{eq1}) with ${\epsilon_1}=1$ between $x=-\infty$ and $x=x_0$,
where $\lambda_1$ is decreasing down the $\lambda_1=0$ axis, whereas it is the solution of (\ref{eq1}) with ${\epsilon_1}=0$ between $x=x_0$ and $x=\infty$,
where $\lambda_1$ increases up the vacuum value $\bar\sigma^2$. This kind of domain wall asymptotically connects the $v^1$ and $v^2$ vacua on the Northern Hemisphere and the $v^3$ and $v^4$ vacua on the Southern Hemisphere.
\FIGURE{\includegraphics[height=4.5cm]{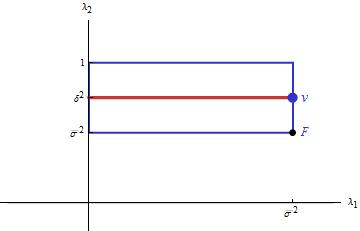}\  \qquad \includegraphics[height=4cm]{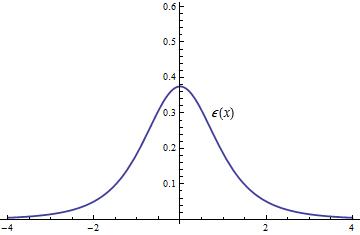}
\caption{The TW orbit in the $P_2$ rectangle displayed as a solid red line (left) and the TW tension density (right) }}

The tension of these domain walls, however, saturates the Bogomolnyi bound {\footnote{Although it seems that the tension of these walls depends not only on the value of $W$ at the vacuum but also on the value of $W(0,\delta^2)$ at $x_0$, (\ref{tenbk1}) is a topological bound. The wall orbit flow is not undefined at $(\lambda_1=0, \lambda_2=\delta^2)$. This point will be clarified further in Section \S. 5 .}}:
\begin{equation}
\Omega({\rm TW})=\frac{2}{\alpha_1^2-\alpha_3^2} \left|W(\bar{\sigma}^2,\delta^2)-W(0,\delta^2)\right|=
\frac{R^3}{\alpha_1^2-\alpha_3^2}\left(\sigma\bar\sigma- (1-2\delta^2) \arccos\sigma \right) \   . \label{tenbk1}
\end{equation}

The inverse image of (\ref{tk3sc}) in ${\mathbb S}^2$
\begin{eqnarray}
\phi_1^{\rm TW}(x;\kappa_1)&=& \frac{(-1)^{\kappa_1} R\delta \sinh \left[ R\sigma\bar{\sigma} (x-x_0)\right]}{\sqrt{\cosh^2 \left[ R\sigma\bar{\sigma} (x-x_0)\right]-\bar{\sigma}^2}}\, \, \, , \hspace{0.2cm}   \phi_2^{\rm TW}(x;\kappa_2)= \frac{(-1)^{\kappa_2} R\sqrt{\delta^2-\bar{\sigma}^2}}{\sqrt{\cosh^2 \left[ R\sigma\bar{\sigma} (x-x_0)\right]-\bar{\sigma}^2}} \nonumber \\  \phi_3^{\rm TW}(x;\kappa_3)&=& \frac{(-1)^{\kappa_3} R\bar{\delta}\cosh\left[ R\sigma\bar{\sigma} (x-x_0)\right]}{\sqrt{\cosh^2 \left[ R\sigma\bar{\sigma} (x-x_0)\right]-\bar{\sigma}^2}}\, \, \, , \quad \kappa_1,\kappa_2,\kappa_3=0,1 \label{tk3cc}
\end{eqnarray}
helps to elucidate the character of these eight new topological walls, see Figure 9. We stress that: (1)
The three field components are different from zero. (2) If $\kappa_3=0$, the topological wall belongs to ${\cal C}_{12}$
($\kappa_1=1$) or to ${\cal C}_{21}$ ($\kappa_1=0$). (3) If $\kappa_3=1$, the topological wall belongs to ${\cal C}_{43}$
($\kappa_1=1$) or to ${\cal C}_{34}$ ($\kappa_1=0$). (4) The sign of $(-1)^{\kappa_2}$ determines the face in ${\mathbb S}^2$ chosen by the wall orbit.

\FIGURE{\includegraphics[height=3.5cm]{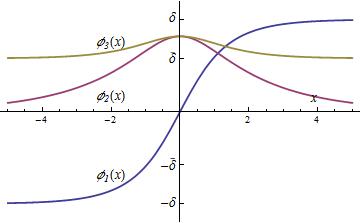}\  \qquad    \includegraphics[height=4cm]{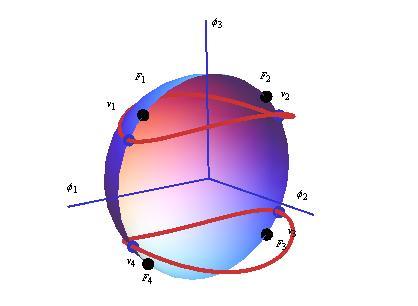}
\caption{Topological wall profiles (\ref{tk3cc})  (left) The orbits (right).}}

One can check that $\Omega({\rm PMW})>\Omega({\rm TW})$ if $\delta>\bar{\sigma}$. Therefore, the stable
topological walls in these sectors are these latter ones (with $\phi_2\neq 0$). This statement is also
clear if one compares the tension density shown in Figure 8 (right) with the density of the other walls living in the
same sector, see Figure 10.

\FIGURE{\includegraphics[height=4cm]{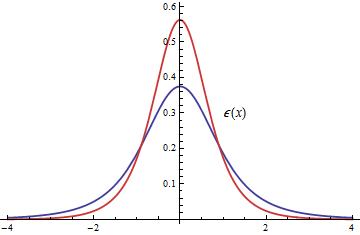}
\caption{Comparison between  the tension densities of the two types of walls in the same topological sector.}}

A better understanding of this point comes from the consideration of the wall orbits in the meridian $\phi_2=0$
in sphero-conical coordinates. The walls in the tropical zone $\lambda_2>\delta^2$(in the ${\cal C}_{14}$ and ${\cal C}_{23}$ sectors) correspond to the orbit: $\lambda_1= \bar{\sigma}^2$. The analytic solution of (\ref{eq2}) is:
\begin{equation}
\lambda_1^{\rm TMW}(x)= \bar{\sigma}^2 \hspace{0.3cm}, \hspace{0.3cm} \lambda_2^{\rm TMW}(x) = \frac{\delta^2 \cosh^2 \left[  R \delta \bar{\delta} (x-x_0)\right]}{\cosh^2 \left[  R \delta \bar{\delta} (x-x_0)\right]-\bar{\delta}^2} \label{tk2t} \qquad .
\end{equation}

Again, (\ref{tk2t}) is the solution of (\ref{eq2}) if $\epsilon_2=0$ between $x=-\infty$ and $x=x_0$, when $\lambda_2$ increases. Between $x=x_0$ and $x=\infty$ the choice in (\ref{eq2}) must be $\epsilon_2=1$ because, then, $\lambda_2$ decreases.

The orbit starts at the vacuum, reaches the equator vertically and bounces back to the vacuum in the sphero-conical plane, see Figure 11(left). The inverse image of this orbit gives the four topological walls in ${\cal C}_{14}$, ${\cal C}_{41}$, ${\cal C}_{32}$,
${\cal C}_{23}$, Figure 11(right). The wall tension (already given in sub-Section \S. 3.1) for this kind of solution can be computed in this context as follows:
\begin{equation}
\Omega({\rm TMW}) = \frac{2}{\alpha_1^2-\alpha_3^2}\left|W((\bar\sigma^2,\delta^2)-W(\bar\sigma^2,1)\right|=
\frac{R^3}{\alpha_1^2-\alpha_3^2}\left[\delta\bar\delta+(1-2\delta^2)\arccos \delta \right] \quad ,\label{tenbk2}
\end{equation}
and the same cautionary remarks concerning (\ref{tenbk1}) are applicable to (\ref{tenbk2}).

In the polar zone $\lambda_2<\delta^2$ we can reproduce the PMW domain wall as a piecewise solution in the sphero-conical plane as follows. There is a solution of (\ref{eq2}) on the orbit $\lambda_1=\bar\sigma^2$ (which is solution of (\ref{eq1})):
\begin{equation}
\lambda_1^{\rm PMW_{VF}}=\bar\sigma^2 \hspace{0.3cm} , \hspace{0.3cm} \lambda_2^{\rm PMW_{VF}}(x) = \frac{\delta^2 \sinh^2 \left[ R \delta \bar{\delta} (x-x_0)\right]}{\cosh^2 \left[  R \delta \bar{\delta} (x-x_0)\right]-\delta^2} \qquad . \label{htk2p1}
\end{equation}
The orbit also starts from the vacuum and reaches the focus F at the point $x_f$ determined by:
\[
x_f\, =\, x_0+\frac{1}{R\delta\bar{\delta}} \, \arcsinh \frac{\bar{\sigma}\bar{\delta}}{\sqrt{\delta^2-\bar{\sigma}^2}}
\]
\FIGURE{\includegraphics[height=4cm]{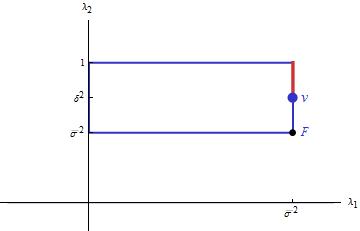}\qquad \includegraphics[height=4cm]{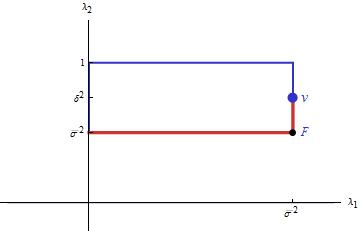}
\caption{The orbits of the TMW (\ref{kinks2}) domain walls (left) and the two-stage orbit of the PMW (\ref{kinks1}) domain walls (right) displayed in $P_2$ as a solid red line.}}

\noindent Thus, (\ref{htk2p1}) is the solution of (\ref{eq2}), with $\epsilon_2=1$ between $x=-\infty$ and $x=-(x_f-x_0)$, whereas
$\epsilon_2=0$ must be chosen between $x=x_f-x_0$ and $x=+\infty$. After hitting the focus for the first time, the orbit changes to proceed along the $\lambda_2 = \bar{\sigma}^2$ axis. The analytic solution of (\ref{eq1})
\begin{equation}
\lambda_1^{\rm PMW_{FP}}(x) = \frac{\delta^2 \sinh^2 \left[ R \delta \bar{\delta} (x-x_0)\right]}{\cosh^2 \left[ R \delta \bar{\delta} (x-x_0)\right]-\delta^2} \hspace{0.3cm} , \hspace{0.3cm} \lambda_2^{\rm PMW_{FP}}(x) = \bar{\sigma}^2 \label{htk2p2}
\end{equation}
departs from the focus at $x=-(x_f-x_0)$, solving (\ref{eq1}) with $\epsilon_1=1$, rebounds at the pole $P$ when $x=x_0$, and travels back with $\epsilon_1=0$ to reach the focus at $x=x_f-x_0$. There is accordingly a continuous sewing of the two stages of the orbit at the focal point, see Figure 11 (left).

The inverse map from (\ref{htk2p1})-(\ref{htk2p2}) to the original field variables produces the four
topological walls shown in Figure 11 (right) that belong to the sectors ${\cal C}_{12}$, ${\cal C}_{21}$,
${\cal C}_{34}$, and ${\cal C}_{43}$ of the configuration space. This continuous gluing of two solutions of two different first-order equation at a point where the flow is undefined  is the bona fide solution of the second-order equations, already found in the sub-Section \S. 2.2 by direct integration. The tension of these walls must be computed in two steps (confirming the result in \S 2.2):
\begin{eqnarray*}
\Omega({\rm PMW})&=& \Omega({\rm PMW_{VF}})+\Omega({\rm PMW_{FP}}) = \\ &=&\frac{2}{\alpha_1^2-\alpha_3^2}\left( \left|W(\bar\sigma^2,\delta^2)-W(\bar\sigma^2,\bar\sigma^2)\right|+
\left|W(\bar\sigma^2,\bar\sigma^2)-W(0,\bar\sigma^2)\right|\right) \\ &=&\frac{R^3}{\alpha_1^2-\alpha_3^2}\left(\delta\bar\delta
-(1-2\delta^2)\arccos \bar\delta\right) \quad .
\end{eqnarray*}

\section{Composite non-topological and topological domain walls}

\subsection{Degenerate families of polar zone non-topological domain walls}

In order to search for more general domain wall solutions we first consider the polar zone:
$\lambda_2\in (\bar\sigma^2,\delta^2)$. The first-order equations (\ref{eq1})-(\ref{eq2}) written in differential form:
\begin{eqnarray}
&& \frac{ (-1)^{\epsilon_1} d \lambda_1 }{2 R (\bar{\sigma}^2
-\lambda_1)(\delta^2-\lambda_1) \sqrt{\lambda_1(1-\lambda_1)}}
+\frac{ (-1)^{\epsilon_2} d \lambda_2 }{2 R (\bar{\sigma}^2
-\lambda_2)(\delta^2-\lambda_2) \sqrt{\lambda_2(1-\lambda_2)}} = 0 \label{traj}\\
&& \frac{ (-1)^{\epsilon_1} \lambda_1 d \lambda_1 }{2 R
(\bar{\sigma}^2 -\lambda_1)(\delta^2-\lambda_1)
\sqrt{\lambda_1(1-\lambda_1)}} +\frac{ (-1)^{\epsilon_2} \lambda_2 d \lambda_2
}{2 R (\bar{\sigma}^2 -\lambda_2)(\delta^2-\lambda_2)
\sqrt{\lambda_2(1-\lambda_2)}} = - d x \label{tstraj}
\end{eqnarray}
lead to the equation for the mechanical orbit -integrating (\ref{traj})- and the rule for the mechanical time
schedule -integrating (\ref{tstraj})- . Therefore, we obtain a two-parametric (the two integration constants
$C_0,C_1\in{\mathbb R}$)
family of domain wall solutions in the implicit form:
\begin{eqnarray*}
&&\int\, \frac{ (-1)^{\epsilon_1} d \lambda_1 }{ (\bar{\sigma}^2
-\lambda_1)(\delta^2-\lambda_1) \sqrt{\lambda_1(1-\lambda_1)}}
+\int \, \frac{ (-1)^{\epsilon_2} d \lambda_2 }{ (\bar{\sigma}^2
-\lambda_2)(\delta^2-\lambda_2) \sqrt{\lambda_2(1-\lambda_2)}} = 2R \, C_1 \\
&&\int  \frac{ (-1)^{\epsilon_1} \lambda_1 d \lambda_1 }{
(\bar{\sigma}^2 -\lambda_1)(\delta^2-\lambda_1)
\sqrt{\lambda_1(1-\lambda_1)}} +\int  \frac{ (-1)^{\epsilon_2} \lambda_2 d \lambda_2
}{ (\bar{\sigma}^2 -\lambda_2)(\delta^2-\lambda_2)
\sqrt{\lambda_2(1-\lambda_2)}} = -2R (x-C_0) \, .
\end{eqnarray*}
To achieve explicit expressions we instead perform the Euler change of variables:
\[
 s_1 = \sqrt{\frac{1-\lambda_1}{\lambda_1}} \quad , \quad    s_2 = \sqrt{\frac{1-\lambda_2}{\lambda_2}}\quad;\qquad 0 < \sigma_2^2 =\frac{\bar{\delta}^2}{\delta^2} < s_2^2 < \sigma_1^2=\frac{\sigma^2}{\bar{\sigma}^2} < s_1^2 < +\infty
\]
Note the inequalities bounding the new variables $s_i, i=1,2$, in terms of the old parameters $\bar\sigma$ and $\delta$ as separation constants. A cosmetic change of notation in favor of $\sigma_1, \sigma_2$ is also introduced to make the formulas more symmetric. The quadratures (\ref{traj})-(\ref{tstraj}) are rationalized:
\begin{equation}
 \frac{1}{\delta^2 \bar{\sigma}^2 R} \sum_{i=1}^2
\int  \frac{(-1)^{\epsilon_i} (1+s_i^2)\ d s_i}{ (\sigma_1^2- s_i^2)
(\sigma_2^2 - s_i^2)} = C_1  \, \, \, , \quad
\frac{1}{\delta^2 \bar{\sigma}^2 R} \sum_{i=1}^2
 \int  \frac{(-1)^{\epsilon_i}  d s_i}{ (\sigma_1^2- s_i^2) (\sigma_2^2 -
s_i^2)} =  x-C_0 \quad . \label{raqua}
\end{equation}
Plugging the simple fraction decompositions
\begin{eqnarray*}
 \frac{1}{(\sigma_1^2-s_i^2)(\sigma_2^2-s_i^2)}&=&\frac{\delta^2\bar\sigma^2}{\bar\sigma^2-\delta^2}
 \left(\frac{1}{\sigma_1^2-s_i^2}-
\frac{1}{\sigma_2^2-s_i^2}\right) \\ \frac{1+s_i^2}{(\sigma_1^2-s_i^2)(\sigma_2^2-s_i^2)}&=&\frac{1}{\bar\sigma^2-\delta^2}
 \left(\frac{\delta^2}{\sigma_1^2-s_i^2}-
\frac{\bar\sigma^2}{\sigma_2^2-s_i^2}\right)
\end{eqnarray*}
in (\ref{raqua}) further simplifies the quadratures:
\begin{equation}
\sum_{i=1}^2 \int \frac{(-1)^{\epsilon_i}  d s_i}{ \sigma_2^2 - s_i^2} =
\delta^2\, R\, (x-C_0-\bar{\sigma}^2\, C_1) \, ,\quad \sum_{i=1}^2  \int \frac{(-1)^{\epsilon_i} d s_i}{ \sigma_1^2 - s_i^2} =
\bar{\sigma}^2\, R\, (x-C_0-\delta^2\, C_1) \qquad . \label{ratquas}
\end{equation}
Integration of the ODE's (\ref{ratquas}) provides the two-parametric family of domain wall solutions:
\begin{equation}
 \arccoth \frac{s_1}{\sigma_2} + \arccoth
\frac{s_2}{\sigma_2} =
R\, \delta\, \bar{\delta}\, (x - x_0) \, \, \, , \quad
 \arccoth \frac{s_1}{\sigma_1} +  \arctanh
\frac{s_2}{\sigma_1} = R\, \sigma\, \bar{\sigma}\, (x-x_0
+ \zeta) \label{hypsol} \quad ,
\end{equation}
now in terms of other two integration constants: $x_0=C_0+\bar{\sigma}^2\, C_1$, $\zeta= (\bar{\sigma}^2-\delta^2)C_1$, $x_0,\zeta \in{\mathbb R}$. Setting the value of the integration constant $x_0$ fixes the ``center of mass" of the wall, whereas different values of $\zeta$ determine the different wall orbits uniquely. We have re-defined the $s_i$ variables in the form $s_1\to (-1)^{\epsilon_1} s_1$, $s_2\to (-1)^{\epsilon_2} s_2$, taking advantage of the parity properties of the inverse hyperbolic functions.

The addition formulas for the hyperbolic functions:
\[
\tanh \left( \arccoth p +\arccoth q \right) \, =\, \frac{p+q}{1+pq}=\frac{1}{
 \tanh \left( \arccoth p +\arctanh q
\right)}
\]
allow us to invert (\ref{hypsol}) and we find:
\begin{equation}
\frac{\sigma_2^2 (s_1+s_2)}{\sigma_2 (\sigma_2^2+s_1 s_2)} = \tanh \left[ R\, \delta\, \bar{\delta} (x-x_0) \right]  \equiv
t_1
\, \, \, , \quad
\frac{\sigma_1 (\sigma_1^2 + s_1 s_2)}{\sigma_1^2 (s_1 + s_2)} =
\tanh \left[ R\, \sigma\, \bar{\sigma} (x-x_0+\zeta) \right]
\equiv t_2 \label{linsys} \quad .
\end{equation}
This is an algebraic linear system in the \lq \lq Vieta variables"
$A= s_1+s_2$\, ,\, $B= s_1s_2$:
\[
\sigma_2 A - t_1 B = t_1 \sigma_2^2 \, \, \, ,
\quad
\sigma_1 t_2 A - B = \sigma_1^2 \quad ,
\]
solvable by means of Cramer's rule:
\[
A(x;x_0,\zeta) = \frac{(\delta^2-\bar{\sigma}^2) t_1}{\delta \bar{\sigma} \left( \sigma \delta t_1t_2-\bar{\sigma}\bar{\delta}\right)} \quad , \quad B(x;x_0,\zeta) = \frac{ \sigma \bar{\delta} \left(\sigma\delta-\bar{\delta}\bar{\sigma} t_1t_2\right)}{\bar{\sigma} \delta \left( \sigma \delta t_1t_2-\bar{\sigma}\bar{\delta}\right)}\qquad .
 \]
$s_1,s_2$ are by definition the roots of the quadratic equation $s^2-As+B=0$:
\[
s_1(x) = \frac{A(x) + \sqrt{A^2(x)- 4 B(x)}}{2} \, , \quad   s_2(x) =
\frac{A(x) - \sqrt{A(x)^2- 4 B(x)}}{2} \quad ,
\]
and thus we find explicit expressions for the family of domain wall solutions
of (\ref{eq1}) and (\ref{eq2}) in the polar zones:
\begin{equation}
\lambda_1^{\rm PZW}(x;x_0,\zeta) = \frac{1}{1+s_1^2(x;x_0,\zeta)} \quad, \quad
\lambda_2^{\rm PZW}(x;x_0,\zeta)= \frac{1}{1+s_2^2(x;x_0,\zeta)} \label{ntdw} \qquad ,
\end{equation}

\FIGURE{\includegraphics[height=3.5cm]{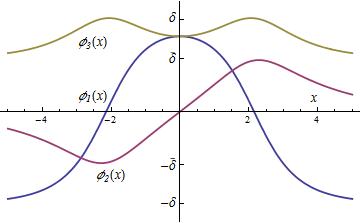}\qquad \includegraphics[height=3.5cm]{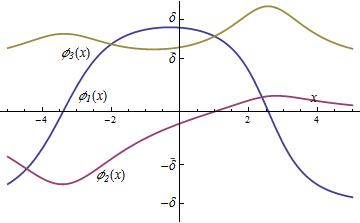}
\caption{Graphics of the domain wall components (\ref{ntk1}, \ref{ntk2}, \ref{ntk3}) for: (left) $x_0=\zeta=0$, and (right) $x_0=1$, $\zeta=3$.}}

\noindent although we stress that in the formula (\ref{ntdw}) the signs of (\ref{eq1}) and (\ref{eq2}) must be chosen to fit with the stages where $\lambda_1$ and $\lambda_2$ are respectively increasing or decreasing.

We return to Cartesian coordinates in ${\mathbb R}^3$ using formula (\ref{spcf}). $\phi_1^2$, $\phi_2^2$ and $\phi_3^2$
are given in terms of $s_1^2+s_2^2$ and $s_1^2s_2^2$. The analytic expressions depending on the spatial coordinate $x$, and the integration constants $\gamma_1$ and $\gamma_2$, are:
\begin{eqnarray}
\phi_1^{\rm PZW}(x;x_0,\zeta)&=& \frac{(-1)^{\kappa_1}R\delta \left( \bar{\delta} \bar{\sigma} -\delta \sigma t_1t_2\right)}{\sqrt{(\delta^2-\bar{\sigma}^2)^2 \, t_1^2+\sigma^2 \bar{\sigma}^2 t_1^2t_2^2-2 \delta \bar{\delta} \sigma\bar{\sigma} t_1t_2+\delta^2\bar{\delta}^2}}\label{ntk1}\\
\phi_2^{\rm PZW}(x;x_0,\zeta)&=& \frac{(-1)^{\kappa_2}R(\delta^2-\bar{\sigma}^2) t_1 \sqrt{1-t_2^2}}{\sqrt{(\delta^2-\bar{\sigma}^2)^2 \, t_1^2+\sigma^2 \bar{\sigma}^2 t_1^2t_2^2-2 \delta \bar{\delta} \sigma\bar{\sigma} t_1t_2+\delta^2\bar{\delta}^2}}\label{ntk2}\\
\phi_3^{\rm PZW}(x;x_0,\zeta)&=& \frac{(-1)^{\kappa_3}R\bar{\delta} (\delta \sigma-\bar{\delta} \bar{\sigma}t_1t_2)}{\sqrt{(\delta^2-\bar{\sigma}^2)^2 \, t_1^2+\sigma^2 \bar{\sigma}^2 t_1^2t_2^2-2 \delta \bar{\delta} \sigma\bar{\sigma} t_1t_2+\delta^2\bar{\delta}^2}}\label{ntk3} \quad .
\end{eqnarray}
The meaning of the integration constants $x_0$, $\zeta$ is now clear:

\medskip

\noindent 1. Besides fixing the center of mass of the composite wall, $x_0$ sets the point $x_f=x_0$, where the field profiles touch the foci. At this point $t_1=0$ and :
    \[
\hspace{-0.5cm} (\phi_1^{\rm PZW})^2(x_0;x_0,\zeta)=R^2\bar{\sigma}^2 \, \, , \, \, (\phi_2^{\rm PZW})^2(x_0;x_0,\zeta)=0 \, \, , \, \, (\phi_3^{\rm PZW})^2(x_0;x_0,\zeta)=R^2\sigma^2 \, \,  .
    \]

\noindent 2. $\zeta$ determines the orbit of the non-topological wall and fixes the relative coordinate between the two centers of these composite walls.

\medskip
\FIGURE{ \includegraphics[height=5cm]{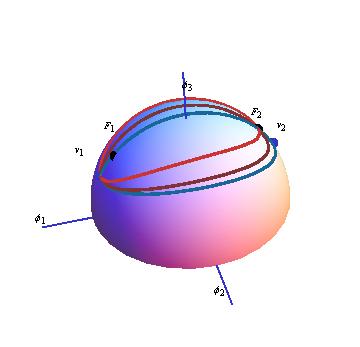}
\caption{Orbits in ${\mathbb S}^2$ for three PZW domain wall solutions: (1) $(\gamma_1=0,\gamma_2=0)$, red. (2) $(\gamma_1=-1,\gamma_2=1)$, brown. (3) $(\gamma_1=1,\gamma_2=6)$, blue.
 (right).}}

In Figure 12 the graphics of the three components of the non-topological domain wall field profiles for two sets of values of $(x_0, \zeta)$ are shown: $\phi_1^{\rm PZW}$ (blue lines) tend to $-\delta$, $\phi_2^{\rm PZW}$ (red lines) tend to $0$, and $\phi_3^{\rm PZW}$ (brown lines) tend to $\bar\delta$ at $x\to +\infty$ and $x\to -\infty$; i.e., the wall solutions go to the same vacuum at both ends of the straight line,which determines the non-topological character of these solutions that belong to the ${\cal C}_{11}$, ${\cal C}_{22}$, ${\cal C}_{33}$,
and ${\cal C}_{44}$ sectors. In the plots of the wall orbits in ${\mathbb S}^2$ drawn in Figure 13,
it is seen that all the orbits start and end at the same vacuum and cross through the opposite focus. More interestingly, the tension densities of these walls depicted in Figure 14 unveil their character as composite
extended objects: they are non-linear superpositions of the PMW and TW
basic walls.
\FIGURE{\includegraphics[height=3cm]{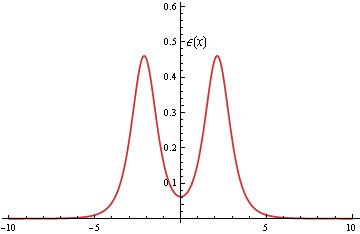}\   \includegraphics[height=3cm]{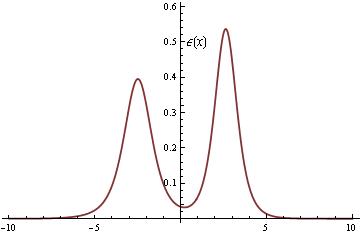}\ \includegraphics[height=3cm]{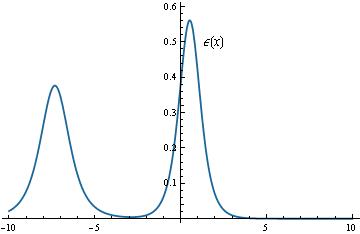}
\caption{Non-topological wall tension densities for:(1) $( x_0=0, \zeta=0)$, red. (2) $(x_0=1,\zeta=2)$, brown. (3) $(x_0=-1,\zeta=5)$, blue.
}}

To further explain these statements, we discuss the orbits in the $P_2$ rectangle. In Figure 15 (left), the plots of three of these wall orbits are shown. The $( x_0=0,\zeta=0)$ orbit (green line) starts at the vacuum, hits the $\lambda_1=0$-axis at the mid-point between $\lambda_2=\delta^2$ and $\lambda_2=\bar\sigma^2$, and runs to the focus $F$. Then, the orbit returns back to the vacuum through the same path
in reversed sense.

The $(x_0=1,\zeta=2)$ orbit (brown line) starts at the vacuum point, hits the $\lambda_1=0$-axis at a point below $\lambda_2=\frac{\delta^2-\bar\sigma^2}{2}$, and run to the focus. Then, the
orbit runs again to intersect the $\lambda_1=0$-axis at the symmetric point over $\lambda_2=\frac{\delta^2-\bar\sigma^2}{2}$, and travels back to reach the vacuum.

The $(x_0=-1,\zeta=5)$
orbit (black line) follows a similar pattern in an extreme way. The trajectory starts from the vacuum, running almost parallel to the $\lambda_2=\delta^2$ orbit, hits the $\lambda_1=0$-axis, and returns slightly departing from the same way until, close again to the vacuum, it turns down toward the focus, almost parallel to the $\lambda_1=\bar\sigma^2$-axis. Just at the focus, the orbit turns to the left, again hitting the $\lambda_1=0$-axis
at a point extremely close to the North Pole where it turns back almost parallel to the $\lambda_1=\bar\sigma^2$-axis.

Again approaching the focus, the orbit finally turns up almost vertically to end at the vacuum. Needless to say, the follow-up of the orbits described here can be interpreted in the inverse sense due to the symmetry $x\to -x$ of the
system. A very important point to emphasize is that the focus $F$ is a conjugate point with respect to the vacuum $V$: a point where all the orbits of a congruence starting in $V$ pass through. This will have consequences in the stability of these domain walls, an issue to be analyzed in a forthcoming publication.

The  $(x_0=-1,\zeta=5)$-orbit is particularly interesting: it is very close to the gluing of the two trial orbits in the ${\cal C}_{12}$ and ${\cal C}_{34}$ sectors of the system. This confirms that these new domain wall solutions are nonlinear superposition of two basic walls. The Bogomolnyi trick provides the tension of any wall in this family :

\begin{eqnarray}
&& \Omega({\rm PZW})=\frac{2}{\alpha_1^2-\alpha_3^2}
\left|W(\bar\sigma^2,\delta^2)-W(0,\delta^2)\right| + \nonumber\\ && \hspace{1.7cm} + \frac{2}{\alpha_1^2-\alpha_3^2}\left(\left|W(\bar\sigma^2,\bar\sigma^2)-W(0,\bar\sigma^2)\right|+
\left|W(\bar\sigma^2,\delta^2)-W(\bar\sigma^2,\bar\sigma^2)\right|\right)\nonumber\\ && \hspace{1.7cm}= \frac{R^3}{\alpha_1^2-\alpha_3^2}
\left(\sigma\bar\sigma+\delta\bar\delta+(1-2\delta^2)({\rm arcsin}\sigma-\frac{\pi}{2}-{\rm arcsin}\delta )\right)\label{ntdwp} \\ && \hspace{1.7cm} = \Omega({\rm PMW})+\Omega({\rm TW}) \nonumber \quad .
\end{eqnarray}
The formula (\ref{ntdwp}) is a remarkably result. First, it means that the tension of all the walls in the family is the same. Second, the tension is equal to the sum of the tensions of the two basic walls that live in the same topological sectors. This wall tension sum rule is another confirmation that the polar domain walls are composed of two basic domain walls.

\FIGURE{\includegraphics[height=4.5cm]{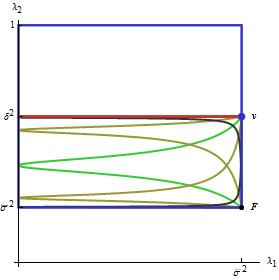}\qquad \includegraphics[height=4.5cm]{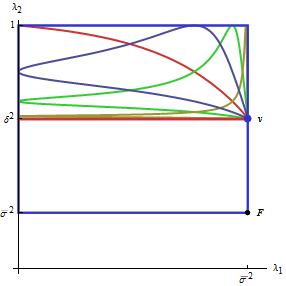}
\caption{Graphics of the domain wall orbits in $P_2$ displayed as solid lines: Polar zone non-topological domain walls (left) and Tropical zone topological walls (right).}}

\subsection{Degenerate families of tropical zone topological domain walls}

If $\delta^2<\lambda_2<1$, the tropical zone, the inequalities satisfied by the $s_i$ variables are:
\[
\sigma_2^2<\sigma_1^2<s_1^2<+\infty \, \, \, , \quad 0<s_2^2<\sigma_2^2  \qquad .
\]
Integration of the ODE's (\ref{ratquas}) gives the domain wall solutions in this zone
\begin{equation}
 \arccoth \frac{s_1}{\sigma_2} + \arctanh
\frac{s_2}{\sigma_2} =
R\, \delta\, \bar{\delta}\, (x -x_0) \, \, \, , \quad
 \arccoth \frac{s_1}{\sigma_1} +  \arctanh
\frac{s_2}{\sigma_1} = R\, \sigma\, \bar{\sigma}\, (x-x_0+\zeta) \label{hypsol1} \quad .
\end{equation}
The only difference with respect to (\ref{hypsol}) is that, $s_2^2$ also being smaller than $\sigma_1^2$,
there are two ${\rm arcth}$ functions entering the solution. The subsequent linear system in Vieta variables
becomes
\[
\sigma_2 t_2A-B=\sigma_2^2 \, \, \, , \quad \sigma_1 t_1 A-B=\sigma_1^2  \quad .
\]
Cramer's rule dictates the solutions:
\[
A(x;x_0,\zeta) = \frac{(\delta^2-\bar{\sigma}^2) }{\delta
 \left( \sigma \bar{\sigma} \delta
t_1-\bar{\delta}\bar{\sigma}^2 t_2\right)} \quad , \quad B(x;x_0,\zeta) =
\frac{ \sigma \left(\sigma\delta \bar{\delta} t_2-\bar{\sigma}
\bar{\delta}^2 t_1\right)}{\delta
 \left( \sigma \bar{\sigma} \delta
t_1-\bar{\delta}\bar{\sigma}^2 t_2 \right) } \quad ,
 \]
\[
s_1(x) = \frac{A(x) + \sqrt{A^2(x)- 4 B(x)}}{2} \, , \quad   s_2(x)
= \frac{A(x) - \sqrt{A(x)^2- 4 B(x)}}{2} \quad ,
\]

\noindent and we have a new two-parametric family of tropical domain walls of equations (\ref{eq1}) and (\ref{eq2})
with the appropriate signs:
\begin{equation}
\lambda_1^{\rm TZW}(x;x_0,\zeta) = \frac{1}{1+s_1^2(x;x_0,\zeta)} \quad, \quad \lambda_2^{\rm TZW}(x;x_0,\zeta)=
\frac{1}{1+s_2^2(x;x_0,\zeta)} \label{ttdw} \qquad .
\end{equation}

\FIGURE{\includegraphics[height=3.5cm]{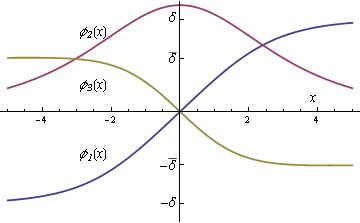}\qquad \includegraphics[height=3.5cm]{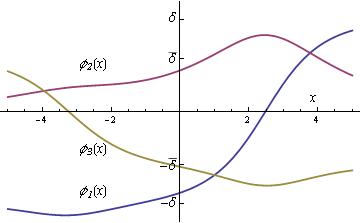}
\caption{Graphics of the tropical zone topological wall components (\ref{tk1}, \ref{tk2}, \ref{tk3}) for: (left) $x_0=0, \zeta=0$, and (right) $x_0=1$, $\zeta=3$.}}

Back in Cartesian coordinates in field space, we find the new family of domain wall solutions:

\begin{eqnarray}
\phi_1^{\rm TZW}(x;x_0,\zeta)&=& \frac{(-1)^{\kappa_1}R\delta \left( \delta \sigma \, t_1-\bar{\sigma}\bar{\delta}\, t_2\right)}{\sqrt{\left( \sigma\bar{\sigma}t_1-\delta\bar{\delta} t_2\right)^2+(\delta^2-\bar{\sigma}^2)^2}}\label{tk1}\\
\phi_2^{\rm TZW}(x;x_0,\zeta)&=& \frac{(-1)^{\kappa_2}R(\delta^2-\bar{\sigma}^2) \, \sqrt{1-t_1^2}}{\sqrt{\left( \sigma\bar{\sigma}t_1-\delta\bar{\delta} t_2\right)^2+(\delta^2-\bar{\sigma}^2)^2}}\label{tk2}\\
\phi_3^{\rm TZW}(x;x_0,\zeta)&=& \frac{(-1)^{\kappa_3}R\bar{\delta} \left( \delta \sigma \, t_2-\bar{\sigma}\bar{\delta}\, t_1\right)}{\sqrt{\left( \sigma\bar{\sigma}t_1-\delta\bar{\delta} t_2\right)^2+(\delta^2-\bar{\sigma}^2)^2}}\label{tk3} \qquad .
\end{eqnarray}
Figure 16 plots the field profiles of two domain walls in this family. They run asymptotically into different
vacua and are thus topological solutions living in topological sectors of the configuration space. In general, the topological wall solutions (\ref{tk1}, \ref{tk2}, \ref{tk3}) belong to the topological sectors: ${\cal C}_{24}/{\cal C}_{42}$
and ${\cal C}_{13}/{\cal C}_{31}$. Their wall tension densities have the form shown in Figure 17, again suggesting the composition of two basic walls: TMW and TW in this case.
\FIGURE{\includegraphics[height=3cm]{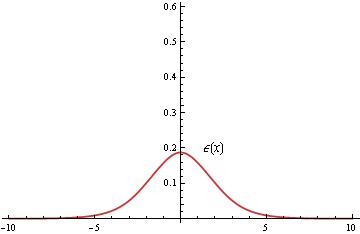}\   \includegraphics[height=3cm]{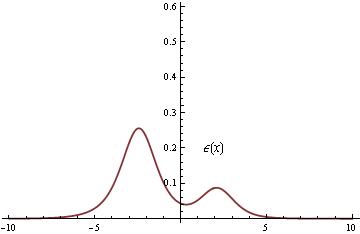}\ \includegraphics[height=3cm]{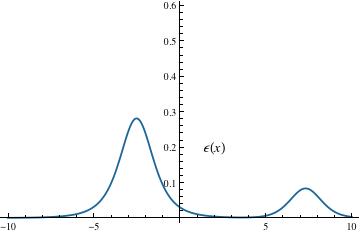}
\caption{Topological wall tension densities for:(1) $( x_0=0,\zeta=0)$, red. (2) $(x_0=-1,\zeta=-2)$, brown. (3) $(x_0=-1,\zeta=-7)$, blue.
}}

The integration constants $x_0$ and $\zeta$ determine the center of mass and relative coordinates of the composite walls, just as in the non-topological polar walls.

\FIGURE{ \includegraphics[height=5.5cm]{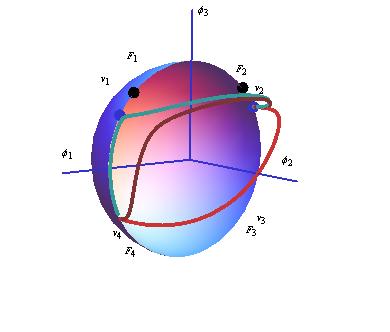}
\caption{Orbits in ${\mathbb S}^2$ of three topological domain wall solutions: (1) $( x_0=0,\zeta=0)$, red. (2) $(x_0=-1,\zeta=-2)$, brown. (3) $(x_0=-1,\zeta=-7)$, blue.}}

The orbits in ${\mathbb S}^2$ of three topological walls are plotted in Figure 18. They join anti-podal vacua and there are no conjugate points in these congruences, a fact that offers a strong hint of stability from the Morse index theorem, see \cite{ItoT}.

Finally, we describe the topological wall orbits in the $P_2$-rectangle in Figure 15 (right). Unlike the non-topological wall orbits that are formed by six, their orbits are composed of four stages. The change of stage takes place when the orbit hits either the $\lambda_2=1$ (the equator) or the $\lambda_1=0$ (the $\phi_1=0$ meridian) edges.
The topological wall orbits do not pass through the foci, the points where the flow is undefined, and therefore
their tension is a true Bogomolny bound, depending only  on the values of the fields at the vacua:

\begin{eqnarray}
\Omega\left({\rm TZW}\right)&=&\frac{2}{\alpha_1^2-\alpha_3^2}
\left(\left|W(\bar\sigma^2,\delta^2)-W(0,\delta^2)\right|+
\left|W(\bar\sigma^2,\delta^2)-W(\bar\sigma^2,1)\right|\right)\nonumber\\ &=& \frac{R^3}{\alpha_1^2-\alpha_3^2}
\left(\sigma\bar\sigma+\delta\bar\delta+(1-2\delta^2)({\rm arcsin}\sigma-{\rm arcsin}\delta )\right)\label{tdwt} \\ &=& \Omega({\rm TMW})+\Omega({\rm TW})\nonumber \quad .
\end{eqnarray}
Accordingly, the topological walls (\ref{tk1}, \ref{tk2}, \ref{tk3}) form a degenerate family of BPS
walls in tension, confirming the stability of these topological defects.

\section{Further comments}

In \cite{AMAJ1}, we discussed the stability properties and quantum features of the domain walls discovered in \cite{AMAJ}. We plan a similar analysis of the domain walls described in this paper in future research.
Nevertheless, we shall briefly comment on these points in this last Section:

\medskip

\noindent {\bf 1.} \underline{Stability}. It is compelling to wonder about the stability of these domain walls. The answer is as follows: \begin{itemize}
    \item The topological basic walls TMW as well as the topological TW (and their anti's) are \underline{stable}.

     \item  The topological basic walls PMW (and their anti's) are \underline{unstable}.

     \item The composite non-topological walls are \underline{unstable}.

     \item The composite topological walls are \underline{stable}.
    \end{itemize}
  The arguments to support these claims are based on: 1) The saturation of Bogomolny bounds, see \cite{Bogomolny}.
  2) The application of the Morse index theorem, see \cite{ItoT}-\cite{MG}. 3) The computation of Jacobi fields, ( not given in this paper), see \cite{AMAJ2}.

\vspace{0.2cm}

\noindent {\bf 2.} \underline{One-loop wall tension shift}. Following the work in \cite{MRNW} on the supersymmetric kink we computed the one-loop mass shift to the only stable kink found in \cite{AMAJ} in Reference \cite{AMAJS}.
    This work has been extended to the massive model with target space ${\mathbb S}^3$ in \cite{AMAJS1}. In this case there are two stable topological kinks but the strategy used was the spectral zeta function regularization developed in the papers \cite{AMAWJ1}-\cite{AMAWJ2}-\cite{AMAWJ3} on linear sigma models with several scalar fields.

\medskip

Finally, a quick remark about what happen if the parameter $\delta$ took other values. First, if
$\delta> 1$ the four ground states become imaginary but the maxima at the intersection between the Equator and
the $\phi_2=0$-meridian become minima. There would be only two vacua on these two points and the structure of the moduli space of domain walls would be very similar to the structure described in \cite{AMAJ}, although the analytic expressions of the domain wall solutions would differ. Second, if $\delta=\bar{\sigma}$ the ground state and the foci would coincide and the two types of wall orbits -connecting either antipodal or non-antipodal pairs of vacuum points- would be topological. All the domain defects would be stable composite topological walls in this case, whereas the basic topological walls would all be stable.

\section{ACKNOWLEDGEMENTS}

We thank to the Spanish Ministerio de Educacion y Ciencia and
Junta de Castilla y Leon for partial financial support under grants
FIS2009-10546 and GR224.

\end{document}